\documentclass[11pt]{article}
\usepackage{geometry}                
\usepackage[parfill]{parskip}    
\usepackage{graphicx}
\usepackage{amssymb}
\usepackage{epstopdf}
\def\Rv{{\mathbf{R}}}
\def\blp{\bigg(}
\def\brp{\bigg)}

\def\fv{{\mathbf{f}}}

\def\Mv{{\mathbf{M}}}

\def\fv{{\mathbf{f}}}

\def\rv{{\mathbf{r}}}

\def\d{\delta}
\def\la{\langle}
\def\ra{\rangle}
\def\Dr{\Delta{\mathbf{r}}}

\def\g{\gamma}
\def\j{\varphi}
\def\l{\lambda}

\def\blp{\bigg (}
\def\brp{\bigg )}
\def\rv{\mathbf{r}}
\def\drv{\Delta \rv}

\def\l{\lambda}
\def\L{\lambda}

\def\wo{\omega_1}
\def\wt{\omega_2}
\def\d{\delta}

\def\F{{\cal{F}}}

\def\a{\alpha}
\def\s{\sigma}
\def\L{\lambda}
\def\k{k}
\def\r{\mathbf{\rho}}

\def\Q{\mbox{{\bf{Q}}}}
\def\Det{\mbox{Det}}
\def\uv{\mathbf{u}}
\def\g{\gamma}
\def\xv{\mathbf{x}}

\def\D{{\cal{D}}}

\def\M{\mathbf{M}}
\def\w{\omega}
\def\co{\cosh(\Lt\w_1)}
\def\ct{\cosh(\Lt\w_2)}
\def\so{\sinh(\Lt \w_1)}
\def\st{\sinh(\Lt\w_2)}
\def\wo{\w_1}
\def\wt{\w_2}
\def\uv{\mathbf{u}}

\def\O{\Omega}

\def\G{{\cal{G}}}

\def\Ds{\Delta s}

\def\H{{\cal{H}}}

\def\Rv{{\mathbf{R}}}

\def\Det{{\mbox{Det}}}
\def\G{{\mathbf{G}}}
\def\j{\varphi}
\def\la{\langle}
\def\ra{\rangle}
\def\fv{\mathbf{f}}

\def\zv{\mathbf{z}}

\def\lt{{l_p}}
\def\Lt{{L}}
\def\e{\epsilon}
\def\av{{\mathbf{a}}}

\author{Greg Morrison$^1$ and D. Thirumalai$^{1,2}$}
\title{Semiflexible Chains in Confined Spaces}

\begin{document}
\maketitle

$^1$Biophysics Program, Institute For Physical Science and Technology, University of Maryland, College Park, Maryland 20742, USA\\
$^2$Department of Chemistry and Biochemsitry, University of Maryland, College Park, Maryland 20742, USA
\\

\abstract{We develop an analytical method for studying the properties of a non-interacting Wormlike Chain (WLC) in confined geometries.  The mean field-like theory replaces the rigid constraints of confinement with average constraints, thus allowing us to develop a tractable method for treating a WLC wrapped on the surface of a sphere, and fully encapsulated within it.  The efficacy of the theory is established by reproducing the exact correlation functions for a WLC confined to the surface of a sphere.  In addition, the coefficients in the free energy are exactly calculated.  We also describe the behavior of a surface-confined chain under external tension that is relevant for single molecule experiments on histone-DNA complexes.  The force-extension curves display spatial oscillations, and the extension of the chain, whose maximum value is bounded by the sphere diameter, scales as $f^{-1}$ at large forces, in contrast to the unconfined chain that approaches the contour length as $f^{-1/2}$.  A WLC encapsulated in a sphere, that is relevant for the study of the viral encapsulation of DNA, can also be treated using the MF approach.  The predictions of the theory for various correlation functions are in excellent agreement with Langevin simulations.  We find that strongly confined chains are highly structured by examining the correlations using a local winding axis.  The predicted pressure of the system is in excellent agreement with simulations but, as is known, is significantly lower than the pressures seen for DNA packaged in viral capsids.}




\section{Introduction}


%


The Wormlike Chain (WLC) model \cite{KP}, that well describes the elasticity of DNA, microtubules, and polyelectrolytes, and is suitable for a polymer with two length scales:  the contour length $L$ and persistence length $l_p$.  When confined to the surface or volume of a sphere, a third length scale appears:  R, the radius of confinement.  The emergence of this new length scale drastically alters the behavior of the WLC, by restricting the conformational space available to the polymer.  A strongly confined WLC will adopt a tightly bent configuration, which is energetically unfavorable in bulk conditions.  Because a number of biologically relevant systems involve stiff chains in confined geometries or adsorbed onto curved surfaces, a general understanding of the WLC model in these geometries is essential.  



The confinement of biopolymers to a curved surface is of interest in a number of systems.  In eukaryotes, the first level of chromosomal compaction of DNA (with $l_p\approx 50$nm) is histone wrapping, with the DNA wrapped around the cylindrical histone (with radius 4.2nm and height 2.4nm) \cite{Histone1,Histone2,HistoneManning}.  The stability of the tightly bent structure is essential in understanding the development of the chromosome.  
In addition, many authors have studied 
the behavior of polymers confined to the surface of both cylinders \cite{OdijkSurface,SurfaceEnergyMin,FJCSurface} and spheres \cite{OdijkSurface,SWSphereSurf,SWSphereSim,MFSurfaceInteract} to discern the free energy scaling and equilibrium behavior of surface confined chains.  In particular, an exact solution for the end-to-end distance $\Rv$ of a WLC confined to the surface of a sphere has been determined \cite{SWSphereSurf}, and confirmed using simulations \cite{SWSphereSim}.

Experiments on the dsDNA-containing bacteriophages $\phi$29 \cite{phiStructure,VirExp11} and $\epsilon$15 \cite{VirExp2}, as well as the $T$ \cite{VirExp1,VirExp3,VirExp5,VirExp8,VirExp10}, P \cite{VirExp4}, and $\l$ \cite{WidomVirus,VirExp6,VirExp7,VirExp9} phage classes,  have determined a number of details of the structures, pressures, and ejection timescales of many viruses.  Regardless of the shape of the viral capsid, it is generally seen that the DNA orders itself in concentric rings \cite{VirExp1,WidomVirus,VirExp2,VirExp3,VirExp5,VirExp6,HudStruct}, with the spacing $\sim$ 0.3nm between rings.  
Single molecule experiments \cite{Buste} have shown the pressure on the capsid walls to be on the order of 60 atm, inducing a significant resistance to the DNA encapsulation.  These observations have generated a number of theoretical studies \cite{HarveyReview}, primarily interested in the packaged structure \cite{TorroidVirus1,TorroidSim,VirusEnergy,MuthVirus,DirectorVirus}, inter-strand spacing \cite{VirusEnergy1,VirusEnergy2,DiscreteVirus,HarveyPhi29}, energy or pressure \cite{OdijkVirus,VirusEnergy1,TorroidVirus1,MuthVirus,ViralTC,BloomfieldEnergy,HelicalSim,ConfScaling}, and the loading or ejection process \cite{VirusEnergy2,EjectionRatchet,TorroidVirus1,TorroidSim,VirusTwistSim,MuthVirus}.  
While the specific geometry of the confining viral capsid varies from phage to phage, the properties of the encapsulated DNA can be studied using spherical \cite{VirusEnergy,VirusTwistSim,TorroidSim} or cylindrical \cite{VirusEnergy1,DiscreteVirus} confinement to a very good approximation.  The study of confined WLCs in these simple geometries is relevant to our understanding of the properties of viruses.  
Interactions between monomers play a significant role in the energetics of structure formation in viral packaging.  However, the initial stages of the encapsulation process may be understood by examining the non-interacting chain, where self-intersections are relatively rare and short range interactions simply renormalize the persistence length \cite{HyeonJCP06}.

%

In order to study the effects of both surface and volume confinement on the behavior of a WLC, we will extend the Mean Field (MF) method \cite{MeanField1,MeanFieldForce,THBook} introduced by Ha and Thirumalai.  
The MF method has been successful in producing tractable theories involving WLCs in many different potentials \cite{MeanFieldForce,CBDT,WLCUnconfInteract}.  In particular, it has been applied to the study of a long, closed WLC on the surface of a sphere \cite{MFSurfaceInteract}.
The paper is organized as follows.  In Sec. \ref{SphereSurfSec}, we extend the MF theory for a WLC on the surface of a sphere of radius $R$, and show that it reproduces all known averages and scaling laws.  We also show that the theory accurately reproduces the correct scaling coefficient of the free energy of confinement.  In Sec. \ref{SurfForceSec}, we adapt the MF theory to study the behavior of a surface-confined WLC subject to an external mechanical force.  
The application of the MF theory to a WLC encapsulated in a sphere (referred to as volume confinement) is discussed in Sec. \ref{SphereVolSec}.  
We show that the volume confined chain is in excellent agreement with simulations using the theory.  As suspected in previous studies, the analytic calculations explicitly show the pressure due to confinement of a non-interacting WLC can not reproduce the large values observed in experiments.   
We also show that the structural order of the confined WLC, that is absent in the bulk, can be understood by using a local winding axis.  The ordering of the chain is purely a consequence of entropic confinement.

\section{Confinement to the Surface of a Sphere}
\label{SphereSurfSec}

{\it{Theoretical Considerations}}\\

\begin{figure}[tpb]
\centering
\includegraphics[width=0.9\textwidth]{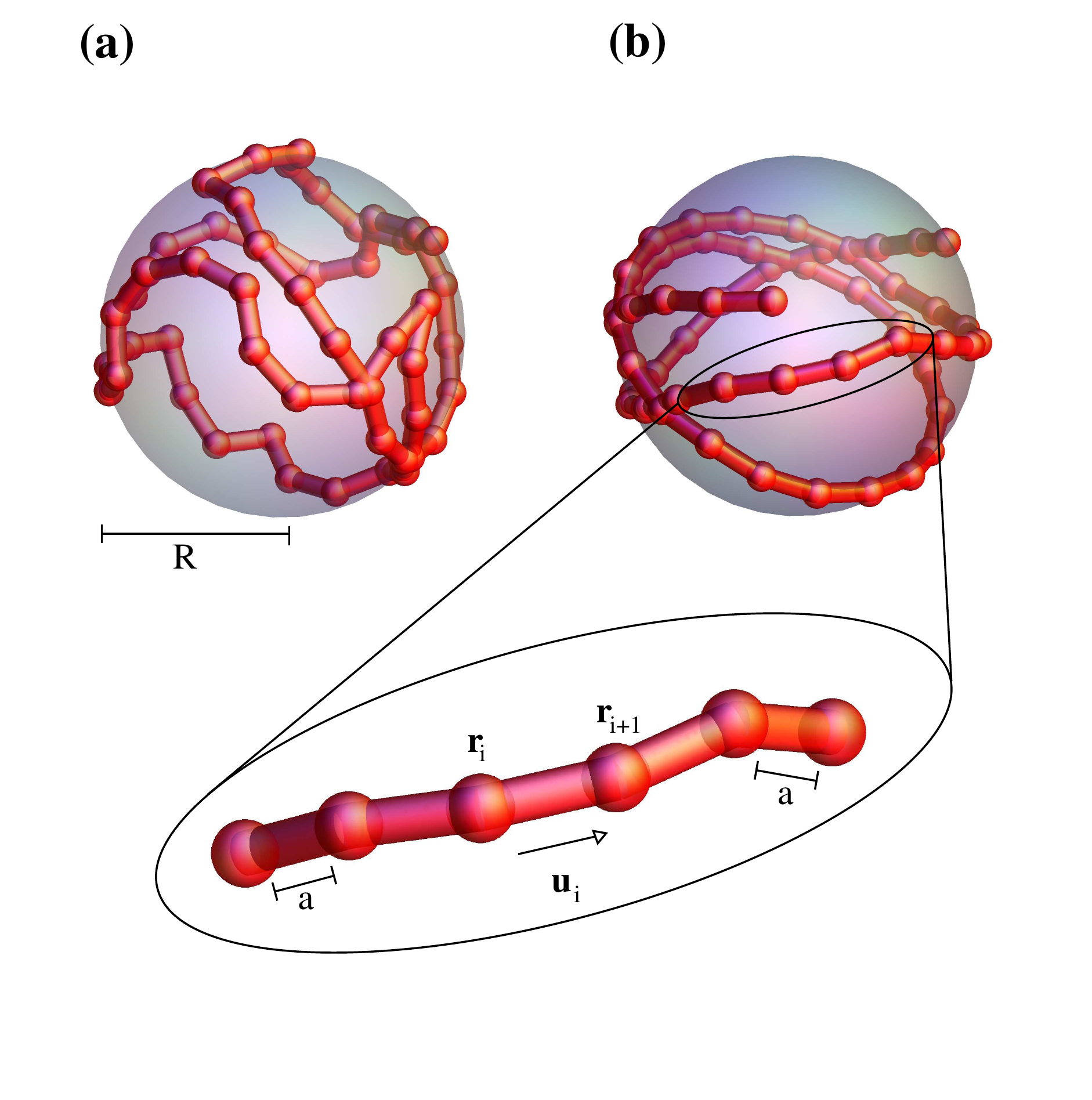}\\
\renewcommand{\baselinestretch}{1}
\small\normalsize
\caption{}
\label{ConfSchematicFig}
\end{figure}

We begin by developing the Mean Field (MF) formalism for a WLC with persistence length $l_p$, fixed inter-monomer spacing $a$, and length $L=Na$, confined to the surface of the sphere of radius $R$.  We define  $\rv_n=(x_n,y_n,z_n)$ the position of the $n^{th}$ monomer, and the bond spacing $\uv_n=\drv_n=\rv_{n+1}-\rv_n$ with $|\uv_n|\equiv a$.  (shown in Fig. \ref{ConfSchematicFig}).   The distribution of the chain in phase space is
\begin{eqnarray}
\Psi_S(\{\rv_n\})&\propto& \prod_n\d(\rv_n^2-R^2)\  \d(\drv_n^2-a^2)\ e^{-l_p/2a^3\,(\drv_{n+1}-\drv_n)^2}\label{discreteeq}\\
&\propto&\int_{-i\infty}^{i\infty}\prod_{n=1}^{N+1} {dk_n d\l_n}\exp\bigg[-\frac{1}{2}al_p\frac{(\drv_{n+1}-\drv_{n})^2}{a^4}-a\l_n\blp\frac{\drv_n^2}{a^2}-1\brp-a k_n\blp\frac{\rv_n^2}{R^2}-1\brp\bigg]\nonumber,
\end{eqnarray}
where the second line follows from the first after a Fourier transform of the delta functions (with the Fourier variables $\{\l_n\}$ and $\{ k_n\}$).  Following Ha and Thirumalai \cite{MeanField1,THBook}, we write the partition function as $Z=\int\prod_n d^3\rv_n\Psi_S(\{\rv_n\})\equiv\int\prod_n d\l_nd k_n$$\exp(-\F_S[\{\l_n,k_n\}])$, which defines the free energy functional for surface confinement, $\F_S$.  The free energy can be written as $\F_S=\F_x+\F_y+\F_z-a\sum_n(\l_n+k_n)$, where $\F_x$ is given by $e^{-\F_x}=\int \prod_n dx_n\ \exp(- \H_x[\{x_n\}])$, with
\begin{eqnarray}
{\H_x}=a\sum_n\blp\frac{l_p}{2}\ \frac{(\Delta x_{n+1}-\Delta x_n)^2}{a^4}+\l_n\frac{\Delta x_n^2}{a^2}+k_n \frac{x_n^2}{R^2}\brp\label{onedconf}.
\end{eqnarray}
We assume $\F_S$ is sharply peaked around a particular set of Fourier variables $\{\l_n,k_n\}=\{\l_n^*,k_n^*\}$, so that $Z\sim \exp(-\F_S^*)$ (i.e. a saddle point approximation).  The optimal values of $\l$ and $k$ are determined by minimizing $\F_S$, i.e. by solving $\partial\F_S/\partial\l_n=\partial\F_S/\partial\k_n=0$.  In this approximation, the Fourier variables play the role of spring constants restricting the position ($\{k_n\}$) and bending ($\{\l_n\}$) of the chain.


Since the discrete Hamiltonian is quadratic in the $x_n$'s, we can write $\F_S=3/2\ \log[\mbox{Det}(\Q)]-a\sum_n(\l_n+k_n)+\mbox{const}$, 
where the symmetric, $(N+1)\times (N+1)$ tridiagonal matrix $\mathbf{Q}$ is given in Appendix A  
(Eqs. \ref{qeq1}-\ref{qeq3}).  The solution for these coupled equations is intractable for large $N$, and additional approximations are necessary to make further progress.  The symmetry of the matrix is respected by the substitution $\l_n\to\L$ and $k_n\to k$, except for exactly three elements near the endpoints (see Appendix A
 for more details).  This is similar to the excess endpoint fluctuation terms found in the unconfined theory \cite{MeanField1,THBook}, where $\l$ was shown to be constant except at the endpoints.  With these observations, we take
\begin{eqnarray}
k_1=k_{N+1}&=&k+\frac{\g_1}{a}-\frac{\g_2}{R}\nonumber\\
k_2=k_{N}&=&k+\frac{\g_2}{R}\label{MFepReplacement}\\
\l_1=\l_{N}&=&\l+\frac{\d}{a}-\frac{a \g_2^2}{R^2}\nonumber,
\end{eqnarray}
with $k_n=k$ and $\l_n=\l$ for all other values of $n$.
The specific forms of the endpoint terms in Eq. \ref{MFepReplacement} are chosen to ensure convergence of the continuum limit.  
Substitution of these values into the Hamiltonian in Eq. $\ref{onedconf}$ and taking the continuum limit (with $a\to0$, $N\to\infty$, and $N a\to L$), we can separate the Hamiltonian into interior and endpoint terms, $\H_x=\H_0+\H_e$, with
\begin{eqnarray}
\H_0&=&\int_0^{\Lt} ds\,\blp\frac{\lt}{2}\,\ddot x^2(s)+\L\,\dot x^2(s)+\k\,\frac{x^2(s)}{R^2}\brp\label{Hcont}\\
\H_e&=&\d(u_0^2+u_L^2)+\g_1\blp\frac{x_0^2}{R^2}+\frac{x_L^2}{R^2}\brp+2\g_2\blp u_0\frac{x_0}{R}- u_L\frac{x_L}{R}\brp\label{He},
\end{eqnarray}
where we have defined $u_0=\dot x(0)$ and $u_L=\dot x(\Lt)$, with $\dot x=\partial x(s)/\partial s$.  The free energy functional $\F_x$ in the continuum limit becomes 
\begin{eqnarray}
\F_x=-\log\bigg[\int d^4\xv\,\exp(-\H_e)\int \D[x(s)]\exp(-\H_0[x(s)])\bigg],\label{Fx}
\end{eqnarray}
with $\xv=(x_0,x_L,u_0,u_L)$, and the total free energy is 
\begin{eqnarray}
\F_S=\F_x+\F_y+\F_z-\L\Lt-\k\Lt-2\d-2\g_1.\label{FreeEnDef}
\end{eqnarray}
The path integral in Eq. \ref{Fx} can be evaluated exactly \cite{Feyn}, and we find 
\begin{eqnarray}
Z_0(\xv)\equiv\int\D[x(s)]\,\exp(-\H_0[x(s)])=K\exp\blp\xv\cdot \M\xv\brp\label{pathintegral},
\end{eqnarray}
 where $\mathbf{M}$ is a 4$\times$4 matrix;  $\mathbf{M}$ and $K$ are evaluated in Appendix B
 , and given explicitly in Eqs. \ref{Meq} and \ref{Keq} in terms of the two frequencies  
 \begin{eqnarray}
 \w_i=\blp\frac{\l}{l_p}\pm\sqrt{1-\frac{2kl_p}{\l^2}}\ \brp^{\frac{1}{2}}\label{omegaDef},
 \end{eqnarray}
Expressions resulting from the propagator in Eq. \ref{pathintegral} can greatly simplified in the limit of large $L\w_i$, which we refer to as strong confinement (see below).

The total free energy functional finally becomes
\begin{eqnarray}
\F_S&=&-3\log\blp\int d^4\xv\ Z(\xv)\brp-\L\Lt-\k\Lt-2\d-2\g_1\nonumber\\
 Z(\xv)&=&Z_0(\xv)\exp({-\H_e})\label{Zdef}
\end{eqnarray}
with $\H_e$ given in Eq. \ref{He} and $Z_0$ in Eq. \ref{pathintegral}.  The optimal parameters $\l$, $k$, $\d$, $\g_1$, and $\g_2$ are obtained by solving the five coupled Mean Field equations, 
\begin{eqnarray}
\frac{\partial\F_S}{\partial\L}=\frac{\partial\F_S}{\partial k}=\frac{\partial\F_S}{\partial\d}=\frac{\partial\F_S}{\partial\g_1}=\frac{\partial\F_S}{\partial\g_2}=0.\label{meanFieldeq}
\end{eqnarray}
Note that, from Eqs. \ref{FreeEnDef} and \ref{meanFieldeq}, the $\L$, $\k$, $\d$ and $\g_1$ derivatives immediately imply, respectively,
\begin{eqnarray}
\frac{1}{L}\int_0^L ds\ \la\uv^2(s)\ra=1,&\qquad& \frac{1}{L}\int_0^L ds\ \la\rv^2(s)\ra=R^2,\nonumber\\
 \la \uv_0^2+\uv_L^2\ra=2,&\qquad&\la\rv_0^2+\rv_L^2\ra=2R^2.\nonumber
\end{eqnarray}
This suggests that the MF approximation is equivalent to replacing the local requirements $\uv^2(s)=1$ and $\rv^2(s)=R^2$ by the global conditions $\la\uv^2(s)\ra=1$ and $\la\rv^2(s)\ra=R^2$.  The parameter $\l$ plays the role of a spring constant that keeps the bond spacing fixed on average, while $k$ is a spring constant that keeps $\la\rv^2\ra=R^2$ on average.  The $\g_2$ derivative in Eq. \ref{meanFieldeq} implies $\la\rv_0\cdot\uv_0\ra-\la\rv_L\cdot\uv_L\ra=0$, as is expected since the rigid constraints require $\uv(s)$ to be tangential to the surface of the sphere (i.e. $\uv(s) \perp \hat\rv$ for all $s$).

The solutions to the mean field equations (Eq. \ref{meanFieldeq}) can be determined exactly for all $L$, $l_p$, and $R$, giving
\begin{eqnarray}
\l=\frac{9}{8l_p}-\frac{l_p}{R^2},\qquad k=\frac{l_p}{2R^2},\qquad \d=\frac{3}{4},\qquad\g_1=\frac{3}{4},\quad\mbox{and  }\g_2=-\frac{l_p}{2R}.\label{SurfSolution}
\end{eqnarray}
We note that $\l$ changes sign for $R^2\le 8l_p^2/9$, because the bonds tend to be more compressed with decreasing $R$, which requires a net repulsion between neighboring monomers to satisfy the constraint $\la\uv^2\ra=1$.   With the solutions to the MF equations in Eq. \ref{SurfSolution}, the frequencies in Eq. \ref{omegaDef} become
\begin{eqnarray}
\w_i=\frac{3}{4l_p}\left( 1\pm \sqrt{1-\frac{16 l_p^2}{9R^2}}\ \right).\label{omegaSurf}
\end{eqnarray}
We note that, in the limit of large $R$, $L\w_1\sim L/l_p$ and $L\w_2\sim Ll_p/R^2$.  Our demarcation of strong confinement, $L\w_i\gg1$, requires long chains ($L\gg l_p$) and sufficiently small radii ($R\ll \sqrt{Ll_p}$).  

%
{\it{Correlation Functions}}

The bending correlation function can be computed directly using the solutions in Eq. \ref{SurfSolution}.  However, in the limit as $R\to \infty$, we find $\la\uv(0)\cdot\uv(L)\ra\to e^{-3L/2l_p}$ as $R\to\infty$, rather than the  expected two-dimensional correlation function, $\la\uv(0)\cdot\uv(L)\ra= e^{-L/2l_p}$.
This suggests that the theory requires a mean field persistence length, $l_0$, much like in the unconfined theory \cite{MeanField1,THBook}.  Substitution of $l_p=3l_0$ 
into the correlation function results in the expected limit as $R\to\infty$.  Ha and Thirumalai, who found a similar result for a three dimensional unconfined WLC with $l_p=3l_0/2$,  argued that the renormalization of $l_p$ in the MF theory arises because of the additional forbidden chain conformations allowed by replacing the $\d$ functions in Eq. $\ref{discreteeq}$ with Gaussians.  Consequently, the mean field persistence length is smaller than the true persistence length.  In the confined theory, we allow three dimensional configurations by replacing the confining $\d$ functions with Gaussians, which would be forbidden by the surface confinement, in addition to relaxing the rigid inter-monomer constraints.  
For this reason, we would expect the confined MF theory to permit additional conformations of the WLC that are forbidden by the rigid constraints, relative to the unconfined theory, thus increasing $l_p/l_0$.
In practice, $l_p$ is often determined by fitting experimental or simulation data to a suitable polymer model.  Hence, the renormalization of $l_p$ within the mean field theory is not a serious concern.   

The correlation functions computed using the MF theory can be written as
\begin{eqnarray}
\la\rv(s)\cdot\rv(s')\ra&=&R^2e^{-|\Ds|/\zeta_S}\bigg[\cosh\blp\frac{|\Ds|}{\zeta_S}\ \O_S \brp+\frac{1}{\O_S}\sinh\blp\frac{|\Ds|}{\zeta_S}\ \O_S\brp\bigg] \nonumber\\
\la\uv(s)\cdot\uv(s')\ra&=&e^{-|\Ds|/\zeta_S}\bigg[\cosh\blp\frac{|\Ds|}{\zeta_S}\ \O_S \brp-\frac{1}{\O_S}\sinh\blp\frac{|\Ds|}{\zeta_S}\ \O_S\brp\bigg]\label{NewCorrelations},
\end{eqnarray}
with $\O_S=\sqrt{1-16l_0^2/R^2}$, and the decay length of the correlations $\zeta_S=1/4l_0$.   We note that Eq. \ref{NewCorrelations} has reproduced the {{exact}} calculation of Spakowitz and Wang \cite{SWSphereSurf}, valid for all values of $L$, $l_0$, and $R$.   The ability to calculate these averages exactly shows the accuracy of the MF method.  We can also verify directly that $\la\uv^2(s)\ra=\la\rv^2(s)\ra/R^2=1$.  Higher order moments are incorrect, though, since $\la\uv^4\ra=\la\rv^4\ra/R^4=5/3\ne 1$ as the rigid constraints would require.

{\it{The Free Energy of Confinement}}

We can determine the free energy of confinement for the system (which does not require the substitution of $l_p=3l_0$) as
\begin{eqnarray}
\beta F\sim \F_S = \frac{9L}{8l_p}+\frac{Ll_p}{2R^2}+\mbox{const}.\label{surfF}
\end{eqnarray}
This result is identical to the scaling predicted by Odijk for a tightly bent WLC \cite{OdijkSurface}.  Additionally, the coefficient of the scaling law agrees with that predicted by Mondescu and Muthukumar \cite{FJCSurface} for the surface-confined Freely Jointed Chain.  
We use the Configurational Bias Monte Carlo (CBMC) method \cite{CBMCBook} to determine the scaling coefficient of the free energy for $L=50a$, for various values of $l_p$ and $R$.  The theoretical curves for $\beta F\sim Ll_p/2R^2+$const are accurate to within $\sim$5\% (see Fig. \ref{SurfF}).

\section{Surface Confined Stiff Chains under Tension}
\label{SurfForceSec}

{\it{Theoretical Considerations}}

The efficacy of the mean field method is in its ability to study the effect of additional potentials in problems involving confined WLC's with relative ease.  In this section, we apply an external tension, $\fv$, to the ends of a surface confined WLC.  For the free chain, the Mean Field method has been shown to give excellent agreement with 
experimental results \cite{BusteDNATension}.  
Such a calculation for the surface-confined WLC will also verify that the mean field method satisfies the confinement on average, even under the extreme situation of a strong pulling force.  

The distribution in phase space of a confined WLC under tension can be written as $\Psi_s(\fv)=\Psi_s\exp[-\beta \fv\cdot(\rv_1-\rv_{N+1})]$, with $\beta=1/k_B T$ and $\Psi_S$ given in Eq. \ref{discreteeq}.  Because the external tension does not generate an energetic term quadratic in the $\rv_n$'s, the MF theory in the previous section can be used with little change.  The discrete free energy functional can be written as $\F_S(f)=\F_x+\F_y+\F_z-\beta f (x_{N+1}-x_1)-a\sum_n(\l_n+k_n)$, with $\F_x$ given in Eq. \ref{onedconf} ($\F_y$ and $\F_z$ are similarly defined), and where we have taken $\fv=f\hat{\mathbf{x}}$.  None of the terms involving $k_n$ or $\l_n$ are altered with the application of the force, and we can again rewrite the quadratic terms of the Hamiltonian using a symmetric, tridiagonal matrix $\Q$ (explicitly given in Appendix 
A
, Eqs. \ref{qeq1}-\ref{qeq3}).  
This again suggests the replacement used in Eq. \ref{MFepReplacement}, with $\l_n$ and $k_n$ constant except near the endpoints.  In the continuum limit ($N\to\infty$, $a\to 0$, and $Na\to L$), we find the free energy
\begin{eqnarray}
e^{-\F_S}
=\blp\int d^4\xv \, Z(\xv)\brp^2\ \blp\int d^4\xv\, Z(\xv)\,e^{-\beta f(x_L-x_0)}\brp\ e^{\L\Lt+\k\Lt+2\d+2\g_1},
\end{eqnarray}
with $Z(\xv)$ given in eq \ref{Zdef} and $\xv=(x_0,x_L,u_0,u_L)$, a result similar to Eq. \ref{Zdef}. The integrals can be evaluated with little difficulty, yielding
\begin{eqnarray}
\F_S(\fv)=\F_S(0)+\frac{(R\beta f)^2}{2}\bigg[\blp\M^{-1}\brp_{11}-\blp\M^{-1}\brp_{12}\bigg]
\end{eqnarray}
where $\F_S(0)$ is the free energy at $f=0$ (Eq. \ref{Zdef}), and $\Mv$ is given in Eq. \ref{Meq}.

Under the assumption that $\Lt\w_i\gg 1$ (a strongly confined chain, see Eq. \ref{omegaDef}), it is not difficult to show that the solutions to the mean field equations (Eq. \ref{meanFieldeq}) become
\begin{eqnarray}
\l=\frac{9}{8l_p}-\frac{l_p}{R^2},\qquad k=\frac{l_p}{2R^2},\qquad\d=\frac{3}{4},\qquad\g_1=\frac{3}{4}\sqrt{1+\frac{4(\beta f R)^2}{9}},\qquad\g_2=-\frac{l_p}{2R}.
\end{eqnarray}
Under the application of a force, only the the endpoints of a strongly confined chain are affected, reflected in the fact that only $\g_1$ depends on $f$.  The interior monomer behavior should be relatively insensitive to $f$ far from the endpoints, so it is not surprising that $\l$ and $k$ are independent of the force.

{\it{Force-Extension Curves}}

The extension as a function of the external tension can be computed using $\la\Rv\ra=-\partial\F_S/\partial (\beta\fv)$.  For a strongly confined chain, we find
\begin{eqnarray}
\frac{\la x_L-x_0\ra}{R}= \frac{1}{3}R\beta f\ \frac{\la\Rv^2\ra_0}{R^2}\blp1+\frac{\la\Rv^2\ra_0}{4R^2}\bigg[1-\sqrt{1+4(R\beta f)^2/9}\bigg]\brp^{-1}\label{ExtensionF},
\end{eqnarray}
with $\la\Rv^2\ra_0=2R^2(1-\la\rv_0\cdot\rv_L\ra_0)$ the average end-to-end distance with $f=0$ (with $\la\rv_0\cdot\rv_L\ra_0$ given in eq. \ref{NewCorrelations}. 
While the force-extension curves for a confined WLC increase monotonically as a function of $f$, the system has rather complicated behavior as a function of $R$.  In Fig. \ref{FEC_SurfFig}(a), we see the extension of a stiff chain ($l_p=150a$, approximately the persistence length of DNA) as a function of $R$ is highly oscillatory for small $R$, due to the non-monotonic behavior of $\la\Rv^2\ra_0$ as a function of $R$.  However, oscillations in $\la\Rv^2\ra_0$ are not observed in more flexible chains, as seen in  \ref{FEC_SurfFig}(b).  This is due to the fact that $\la\rv_0\cdot\rv_L\ra_0\sim e^{-L/4l_0}$, so that the oscillations in $\la\Rv^2\ra_0$ for longer or more flexible chains are damped out.  The non-monotonic behavior observed in  \ref{FEC_SurfFig}(a) is thus due to finite-size effects.

The asymptotic limits of Eq. \ref{ExtensionF} are
\begin{eqnarray}
\la x_L-x_0\ra\sim\left\{\begin{array}{cc}\beta f\la\Rv^2\ra_0/3 & R\beta f\ll 1 \\2R-\frac{3}{\beta f}(1-4\frac{R^2}{\la\Rv^2\ra_0})+O(f^{-2}) & R\beta f\gg 1\end{array}\right.  .
\end{eqnarray}
In the low force regime, the system has the expected linear response to the tension, and $\la x_L-x_0\ra\le 2R$ for all values of the force.  
Surprisingly, though, the scaling of $\la x_L-x_0\ra-2R\sim f^{-1}$ in the high force regime differs from the scaling of the unconfined chain, $\la x_L-x_0\ra-L\sim f^{-1/2}$.  The change in the large-force scaling laws is linked to the fact that only the endpoints are affected by the force for a surface confined chain.  For a free WLC, the extension of the chain comes about by alignment of all bonds with the force axis.  When confined to the surface of the sphere, the extension occurs primarily by to the translation of the endpoints to the poles of the sphere, rather than a global realignment of the bond vectors.  This is reflected in the fact that $\g_1$, which controls the position of the endpoints, is the only mean field variable dependent on $f$.  We note as well that the $f^{-1/2}$ scaling is seen in the MF theory for the free WLC, and comes about due to the fact that $\l$ (which determines the behavior of all of the bonds) becomes a function of $f$.  

It is also possible to numerically solve the mean field equations for small $L/R$ and $R\beta f\gg 1$ (strong stretching limit), where we find $\la x_L-x_0\ra\approx 2R\sin(L/2R)$ for $L\le\pi R$,
 the exact end-to-end distance of a fully stretched chain confined to the surface of a sphere.  The mean field method thus satisfies the confining constraints on an average, even under high forces, and again predicts the lower moments exactly.  

Finally, we can determine the free energy of a confined WLC under tension, in the limit of strong confinement:
\begin{eqnarray}
\beta F=\frac{9L}{8l_p}+\frac{Ll_p}{2R^2}
+3\log\bigg[1+\sqrt{1+\frac{4(\beta f R)^2}{9}}+O(e^{-3L/4l_p})\bigg]-\beta f\la x_L-x_0\ra+\mbox{const} \label{ForceExtEq},
\end{eqnarray}
The force-dependent terms in Eq. \ref{ForceExtEq} are not extensive because the tension only strongly effects the endpoints of the chain.  If we neglect terms on the order of $e^{-3L/4l_p}$, $F$ becomes tension-dominated when $f$ exceeds a critical force $R\beta f_c\sim Ll_p/4R^2$.
Because the only force-scale in the problem is $\beta P A\sim Ll_p/R^2$, with $A$ the surface area of the sphere, the scaling of this critical force is expected.  We expect the leading coefficient to be correct, due to the accuracy of our expression for the free energy of a WLC without the external tension (see Fig. \ref{SurfF}).  

It is amusing to estimate $f_c$ for a strand of DNA wrapped around a histone \cite{Histone1,Histone2,HistoneManning}, with $l_p\approx 50$nm, $L\approx 43$nm, and $R\approx 4$nm.  We find the tension dominates the free energy when $f>f_c\approx34$pN, which is significantly larger than the force required at each unwrapping event seen in single molecule experiments on histones \cite{Histone2}.  However, as it has been observed that the tilting of the histone with respect to the force axis is of great importance when determining the behavior of the system \cite{Histone1}, which the mean field theory does not take into account.  Our result only provides an upper bound on the unravelling force.

\section{Wormlike chains confined to the interior of a sphere}
\label{SphereVolSec}
{\it{Theoretical Considerations:}}\\

The mean field theory for computing the average properties of a surface confined chain can be extended to studying the effects of volume confinement.  The distribution in phase space of a WLC confined to the interior of a sphere is
\begin{eqnarray}
\Psi_V(\{\rv_n\})\propto\prod_n\ \Theta(R^2-\rv_n^2)\ \d(\drv_n^2-a^2)\ e^{-l_p(\drv_{n+1}-\drv_n)^2/a^3}\label{PsiVol}
\end{eqnarray}
where $\Theta(x)$ is the Heaviside step function, that ensures that each monomer is contained within the sphere.  The last two terms in Eq. \ref{PsiVol} are identical to the ones in $\Psi_S$ (Eq. $\ref{discreteeq}$).  The similarities between the two distributions suggest that volume confinement can be treated at the mean field level as well.  
Unfortunately, the $\Theta$ function in Eq. \ref{PsiVol}, that ensures the chain is within the interior of the sphere of radius $R$, can not be dealt with as simply as the $\d$ functions found in Eq. \ref{discreteeq} at the mean field level.  It is not difficult to show that, for a single particle confined within a sphere, simply minimizing the Fourier Transform of the $\Theta$ function does not give the correct value of $\la\rv^2\ra$.  
However, we may formally write
\begin{eqnarray}
\Psi_V\propto\int_{-i\infty}^{i\infty}\prod_n{dk_n d\l_n}\exp\bigg[-\frac{1}{2}al_p\frac{(\drv_{n+1}-\drv_{n})^2}{a^4}-a\l_n\blp\frac{\drv_n^2}{a^2}-1\brp\nonumber\\
-a k_n\frac{\rv_n^2}{R^2}-g(ak_n)\bigg],\label{discreteeqvol}
\end{eqnarray}
where $g$, an undetermined function, is chosen such that free energy minimization satisfies the rigid, local constraints on average ($\uv^2(s)=a^2$ and $\rv^2(s)\le R^2$).  
We immediately see that the same substitution of interior (i.e. $k_n\to k$ and $\l_n\to\l$) and endpoint terms (Eq. \ref{MFepReplacement}) will satisfy the symmetry of $\Q$ (see Appendix A
, and Eqs. \ref{qeq1}-\ref{qeq3}), due to the similarities between Eq. \ref{discreteeq} and Eq. \ref{discreteeqvol}.   This allows the problem of volume confinement in the continuum limit to be written in terms of the mean field variables $\l$, $k$, $\d$, $\g_1$, and $\g_2$, with the free energy expressible as $\F_V=\F_x+\F_y+\F_z-G[\l,k,\d,\g_1,\g_2]$.   $\F_x$ is defined in Eq. \ref{Fx}, and $G$ constrains the minimization of $\F$ (i.e. contains the as yet undetermined Lagrange multipliers).  
%




The treatment of volume confinement at the mean field level is more difficult than the case of surface confinement for a number of reasons.  {In the case of surface confinement, we replaced the strict constraint of $\rv^2(s)\equiv R$ with the global constraint $\frac{1}{L}\int_0^L ds\la\rv^2(s)\ra=R^2$.  
While the average monomer position for a volume confined WLC is not known {\it{a priori}}, we expect that interior monomers, those far from the endpoints, will have a uniform behavior.  This suggests that we write
\begin{eqnarray}
\frac{1}{L}\int_0^L ds{\la\rv^2(s)\ra }{}\equiv \r R^2\label{rhodef},
\end{eqnarray}
for some unknown $\r$, which may depend on $L$, $l_p$, and $R$.
Eq. \ref{rhodef} is equivalent to the requirement at the mean field level 
\begin{eqnarray}
\frac{\partial \F}{\partial k}=L\r
\end{eqnarray}
(see Eq. \ref{Hcont}).  
}
{Unlike the surface case, the average position of the endpoints within the sphere need not be identical to the average position for interior points of the chain, i.e. $\la\rv^2(s)\ra\ne\mbox{const}$ for volume confinement.  At the mean field level, this can be treated approximately by the restriction
\begin{eqnarray}
\la\rv^2_0\ra=\la\rv^2_L\ra=\r_0 R^2\label{r0def},
\end{eqnarray}
with $\r_0\ne \r$ an unknown parameter.  This is implemented using
\begin{eqnarray}
\frac{\partial\F}{\partial\g_1}=2\r_0
\end{eqnarray}
(see Eq. \ref{He}).  
}
{In addition to the nonuniformity at the endpoints, volume confinement allows for different fluctuations in the bending at the endpoints.  Because $\uv(s)$ need not be perpendicular to $\hat\rv$ (as was the case for surface confinement), $\la\uv(s)\cdot\rv(s)\ra\ne0$, which must be accounted for at the mean field level as well.  
Since $\uv(s)=d \rv(s)/d s$ changes sign under the transform $s\to L-s$, it is simple to show that $\la\uv_0\cdot\rv_0\ra=-\la\uv_L\cdot\rv_L\ra$.  In particular, if the endpoints of a confined chain are found near the wall of the sphere, the direction of the bond vectors at the endpoint will be restricted, pointing away from the wall of the sphere and giving $\la\rv_L\cdot\uv_L\ra>0$.  We then restrict
\begin{eqnarray}
{\la\uv_L\cdot\rv_L\ra}{}=-{\la\uv_0\cdot\rv_0\ra}{}\equiv \r_cR\label{rcdef}.
\end{eqnarray}
The unknown parameter $\r_c$ represents the average correlation between the position and the bending at the endpoints of the chain.    
Eq. \ref{rcdef} is expressed at the mean field level as
\begin{eqnarray}
\frac{\partial \F}{\partial \g_2}=-4\r_c
\end{eqnarray}
(see Eq. \ref{He}).   
}
The lagrange multipliers for both $\l$ and $\d$ remain unchanged at the mean field level, with $\partial \F/\partial\l=L$ and $\partial\F/\partial \d=2$ (see Eq. \ref{Zdef}).  The three Mean Field parameters $\r$, $\r_0$ and $\r_c$ can not be computed within the framework of the MF theory, and must be supplied using some other method.  We use low friction Langevin dynamics simulations to determine the equilibrium behavior of a WLC confined to the interior of a sphere.  The details of our simulations are given in Appendix C.

Up to the three undetermined parameters ($\r$, $\r_0$, and $\r_c$), we can write (as we did in Eq. \ref{FreeEnDef}) the mean field free energy for volume confinement as
\begin{eqnarray}
\F_V=\F_x+\F_y+\F_z-\l L-2\d-\r kL-2\r_0\g_1+4\r_c\g_2
\end{eqnarray}
where $\F_x$ is identical to the one dimensional free energy functional for the surface case (eq \ref{Fx}).   
The mean field equations for the volume case, 
given in Eq. \ref{meanFieldeq}, are quite similar to the surface equations, and can be solved in the limit of strong confinement (i.e. $L \w_i\gg 1$, see Eq. \ref{omegaDef}).  We find
\begin{eqnarray}
\l=\frac{9}{8l_p}-\frac{l_p}{\r R^2}\qquad k=\frac{l_p}{2\r^2 R^2}\qquad\d=\frac{3(\r_0+\r_c^2)}{4(\r_0-\r_c^2)}\qquad \nonumber\\
\g_1=-\frac{3}{4\r}+\frac{3}{2(\r_0-\r_c^2)}\quad\qquad\quad\g_2=-\frac{l_p}{2\r R}+\frac{3\r_c}{2(\r_0-\r_c^2)},\label{VolRoots}
\end{eqnarray}
and the frequencies in Eq. \ref{omegaDef} become $\w_i= 3/4l_p(1\pm\sqrt{1-16l_p^2/9\r R^2})$.  The solutions for the $\w_i$'s, which define the average behavior over the entire length of the chain, are identical to those found for a wormlike chain confined to the surface of a sphere of radius $\sqrt{\r}\,R$ (see Eq. \ref{omegaSurf}).  However, the endpoint terms differ from the surface confined system (Eq. \ref{SurfSolution}), allowing for differing behavior between the monomers at the ends and those the interior of the chain.  

In Figure \ref{rmagFig}, we show the simulated average monomer positions as a function of $s$ for varying $R$ and $l_p$.  Fig. \ref{rmagFig}a shows that ``interior'' monomer behavior (where $\la\rv^2(s)\ra\approx\r R^2=\mbox{const}$) begins to emerge in the range $2R\ge s\ge L-2R$. 
Significant deviations from $\la \rv^2(s)\ra\approx \r R^2$ occur near the endpoints of the chain over a range of $s\approx 2R$, due to the differing fluctuations in the endpoint monomers (see Eq. \ref{VolRoots}).  
The range of the endpoint effect makes physical sense:  if $\rv_0$ is near the boundary of the sphere, the bending energy near the endpoint will be lower if $\uv_0$ is directed towards the center of the sphere, as opposed to being directed towards the wall.  This suggests that segments of the chain near the endpoints will be directed inwards, giving rise to the decrease in $\la\rv^2\ra$ seen in Fig. \ref{rmagFig}.  
Endpoint effects will dominate the behavior of the chain until the segment comes into contact with the opposite side of the sphere on average, a distance of at most $2R$.

In Fig. \ref{rmagFig}b, we see that increasing the persistence length of the chain while keeping $R$ fixed changes the values of $\r$ and $\r_0$ (reflected in the overall increase 
in $\la \rv^2\ra$), but does not significantly alter the qualitative behavior of $\la\rv^2\ra$ as a function of $s$.   
For strong confinement, fluctuations in $\la \rv^2\ra$ are small far from the endpoints, clearly indicative of an effective surface confinement on a sphere of radius $\approx \sqrt{\r} R$, consistent with the results of the MF theory.

{\it{Determination of $\r$, $\r_0$, and $\r_c$:}}

We perform a number of simulations in order to determine the mean field parameters $\r$, $\r_0$ and $\r_c$ for varying $l_p$ and $R$, shown in Fig. \ref{rhoplotFig}(a).  We find that, for long chains ($L=100a$ and $L=200a$, there is virtually no variation in any of the mean field parameters with respect to $L$.  Since $\r$ determines the effective surface confinement (see Fig. \ref{rmagFig}b), increasing $L$ does not change $\r$, as the chain simply wraps further around the effective surface at $\sqrt{\r}\ R$.  $\r_0$ and $\r_c$ are likewise independent of the length of the chain, due to the fact that the behavior of the endpoints is only weakly dependent on $L$ as long as $L\gtrsim 2R$.  
Since the only remaining length scales in the system are $l_p$ and $R$, we expect that all of the mean field parameters depend only on the ratio $l_p/R$.   This is confirmed in Fig. \ref{rhoplotFig}(a), as the computed values for $\r$, $\r_0$ and $\r_c$ each collapse onto a single curve as a function of $l_p/R$.
We find for long, stiff chains (with $L/R\gg 1$ and $l_p/R\gg 1$) that $\r\lesssim 0.9$, $\rho_0\lesssim 0.95$, and $\r_c\lesssim 0.25$.  Most systems of biological interest (the viral packing of DNA, for example) are in the strongly confined regime.  

{\it{Correlation Functions:}}


For interior monomers (where the system is confined approximately to a sphere of radius $\sqrt{\r}\ R$, see Fig. \ref{rmagFig}), we find that the bending correlation function 
converges on $\la\uv(s)\cdot\uv(s')\ra\to e^{-3|\Ds|/2l_p}$ in the limit of $R\to\infty$, rather than the expected unconfined limit of $\la\uv(s)\cdot\uv(s')\ra=e^{-|\Ds|/l_p}$.  
This suggests the Mean Field persistence length $l_p=3/2l_0$ for large $R$, identical to the result found in the unconfined MF theory \cite{MeanField1,THBook}.  
However, for $l_p/R\gg 1$, the system is effectively confined to a sphere of radius $\sqrt{\r}\ R$.  As the surface MF persistence length is given by $l_p=3l_0^S$, we expect the volume confined $l_0$ to be a function of $R$.  The ratio $l_p/l_0\equiv \a$ should be independent of $L$ for long chains, since we have seen that the development of near-surface confinement depends only on the ratio $l_p/R$ (Fig. \ref{rhoplotFig}a).  
We expect $3/2\le \a(l_p/R)\le 3$, i.e. a wormlike chain confined to the interior of a sphere will behave somewhere in between a free wormlike chain (3 dimensional), and a surface confined wormlike chain (two dimensional).  
For interior monomers, we find
\begin{eqnarray}
{\la\rv(s)\cdot\rv(s')\ra}{}&\approx&\r R^2 e^{-|\Delta s|/\zeta_V}\bigg[\cosh\blp\frac{|\Delta s|}{\zeta_V}\O_V\brp+\frac{1}{\O_V}\sinh\blp\frac{|\Delta s|}{\zeta_V}\O_V\brp\bigg]\nonumber\\
\la{\uv(s)\cdot\uv(s')}{}\ra&\approx&e^{-|\Delta s|/\zeta_V}\bigg[\cosh\blp\frac{|\Delta s|}{\zeta_V}\O_V\brp-\frac{1}{\O_V}\sinh\blp\frac{|\Delta s|}{\zeta_V}\O_V\brp\bigg]\label{bend_volume},
\end{eqnarray}
with $\zeta_V=4\a l_0/3$ and $\O_V=\sqrt{1-16\a^2l_0^2/9\r R^2}$.  Near the endpoints, the correlation functions become more complicated, due to the dependence of the behavior of the endpoints on $\r_0$ and $\r_c$.

%

In order to determine $l_p/l_0=\a(l_p/R)$, we turn to our simulation results again.  
 The simulated bending correlation function is fit using Eq. \ref{bend_volume}, with $\a$ as the only a fitting parameter, with the resulting values are shown in Fig. \ref{rhoplotFig}b.  We find that $\a$ does indeed vary with only $l_p/R$ for weak confinement, and $\a\gtrsim 3/2$ in this range.  For stronger confinement, the bending correlation function is only weakly dependent on $\a$, with large fluctuations in the fitting parameter for increasing $l_p/R$.   
However, the saturating value appears to be $\a(\infty)\approx 5/2$.   The fact that $\a$ does not reach the maximal value of $l_0/l_p=3$ is not surprising, as volume confinement still allows fluctuations in $\la\rv^2\ra$ forbidden by surface confinement.
The decay length in Eq. \ref{bend_volume}, $\zeta_V=4\a l_0/3\approx 3.3l_0$ for strongly confined chains, 
is strictly less than than the decay length for surface confinement, $\zeta_S=4l_0$, again due to the larger number of configurations that are available to volume confined chains.  
The agreement between simulation and theory is excellent not only for the bending correlation function (Fig. \ref{bendCorrFig}), but also the agreement for the position correlation function is equally as good (data not shown).

{\it{Probes of structures:}}

It is of interest to probe the confinement-induced structure in a WLC, that is a thermally fluctuating filament in the bulk.  Information about the structure of a stiff chain confined to the interior of a sphere can be determined using the local winding axis of the chain.  The unit local winding axis of bonds $i$ and $i+1$ (the axis about which $\uv_i$ and $\uv_{i+1}$ wind) is given by \cite{SWSphereSim} $\hat\av_i=\av_i/|\av_i|$, with
\begin{eqnarray}
\av_i=\uv_{i}\times\uv_{i-1}\ ,
\end{eqnarray}
Analytical work with the local winding axis is difficult, because 
\begin{eqnarray}
\hat\av_i\cdot\hat\av_{i+1}=\frac{\cos(\theta_{i-1,i+1})}{\sin(\theta_{i-1,i})\sin(\theta_{i,i+1})}-\cot(\theta_{i+1,i})\cot(\theta_{i-1,i}) \label{WindingAxisAngles},
\end{eqnarray}
where we have defined $\cos(\theta_{i,j})=\uv_{i}\cdot\uv_j$, giving rise to a four-point correlation function.  The details of this result are shown in Appendix D
.  While directly computing the average of Eq. \ref{WindingAxisAngles} is analytically intractable, the symmetry of the problem shows that, for a free WLC, $\la\hat\av_i\cdot\hat\av_{i+1}\ra=0$ (since $\uv_{i+1}$ may be freely rotated about the $\uv_i$ axis without changing $\theta_{i,i+1}$).  
The simulations show that the local winding axes for interior bonds are highly correlated for strongly confined chains, with $\la\hat \av_i\cdot\hat\av_{i+1}\ra$ collapsing on a single, increasing curve as a function of $al_p/R^2$
(Fig. \ref{windFig}a).  
Correlations in the winding axis will thus develop more readily for smaller radii than will the oscillations seen in the bending correlation function (depending on the ratio $l_p/R$, see Eq. \ref{bend_volume} and Fig. \ref{bendCorrFig}).  
The endpoints of the chain are not strongly correlated to the interior behavior (Fig. \ref{windFig}a, inset), with a precipitous drop to $\la\hat\av_i\cdot\hat\av_{i+1}\ra\lesssim0.1$ at the endpoints.  This sharp drop suggests that the endpoints of the chain behave more like an unconfined chain than do the interior monomers, with $\la\hat\av_i\cdot\hat\av_{i+1}\ra\approx 0$, consistent with our physical picture of the origin of the endpoint effects (see the discussion above).

Correlations between the winding axes for interior monomers as a function of their separation $\Ds$ appear exponentially distributed (Fig. \ref{windFig}b), with a best fit
\begin{eqnarray}
\la\hat\av(s)\cdot\hat\av(s')\ra\approx \la\hat\av_i\cdot\hat\av_{i+1}\ra e^{-|\Ds|/2l_0}\label{axisdecay}.
\end{eqnarray}
The sharp drop in $\la\hat\av(L/2)\cdot\hat\av(s')$ near the endpoints shows that the wrapping near the ends of the chain is uncorrelated to the interior wrapping, consistent with the behavior seen in the inset of Fig. \ref{windFig}a.  Forrey and Muthukumar \cite{MuthVirus} use $\la\av_i\cdot \hat\zv \ra$ as an order parameter in the study of the wrapping of DNA within the $\phi$29 phage (a natural choice, as the DNA loaded into the capsid along the $z$-axis).  
They find weak correlations between the local winding axis and the $z$-axis, as is expected due to the lack of correlations between the interior and the endpoints (Fig. \ref{windFig}).

{\it{Pressure Estimates:}}

The free energy and pressure of the volume confined WLC can be computed using our mean field roots (Eq. \ref{VolRoots}).  The exact expressions are somewhat lengthy, due to the endpoint terms involving $\r_0$ and $\r_c$, but in the limit of small $R$ (relevant for most physical systems) we find
\begin{eqnarray}
\beta F\sim \frac{Ll_p}{2\rho R^2}-\frac{2\r_c l_p}{\rho R}-3\log(R)+\mbox{const}\qquad\qquad\beta P V\sim\frac{Ll_p}{3\r R^2}-\frac{2\r_c l_p}{3\r R}+1\label{press_vol},
\end{eqnarray}
with $P=-\partial F/\partial V$.  The $R^{-1}$ terms in the free energy and pressure are not present in the surface confined case, and are due entirely to the nonuniformity in $\la \rv^2(s)\ra$ as a function of $s$ (see Fig. \ref{rmagFig}), reflected in the fact that this term is proportional to $\r_c$.  
The coefficient of the $R^{-1}$ term in Eq. \ref{press_vol} is negative, due to the fact that portions of the chain near the endpoints will be found on average closer to the center of the sphere than the interior monomers (as seen in Fig \ref{rmagFig}), resulting in a decrease in the pressure.  The excellent agreement between simulations and the theoretical predictions (Fig. \ref{pressFig}) shows that Eq. \ref{press_vol}, with $\r\approx 0.9$, $\r_0\approx 0.95$, and $\r_c\approx 0.25$ for strong confinement, can be used in the calculation of the entropy of confinement for a WLC. 

In order to determine the pressure directly from the simulations, we compute 
\begin{eqnarray}
PA=\sum_i {\mathbf{f}}_{i\to wall}\cdot\hat{\mathbf{r}}
\end{eqnarray}
with $A=4\pi R^2$ the surface area of the sphere, and $ {\mathbf{f}}_{i\to wall}$ the force of the $i^{th}$ monomer on the wall.  In Fig. \ref{pressFig}, we show the simulated results along with the full mean field expression for the pressure  (of which Eq. \ref{press_vol} is the limit of small $R$).  We find the agreement is excellent for a large range of $L$, $l_p$ and $R$, particularly for small $R$ where endpoint effects are less important.

We can compare our results to the experimental pressures determined by Smith et. al \cite{Buste}, using the $\phi29$ virus.  The viral capsid is not spherical, with an icosohedral shell of radius $\approx$21nm and height $\approx$54nm, but has a volume equivalent to a sphere of radius $\approx 26$nm.  The fully packed virus contains a strand of DNA of length 6.6$\mu$m, with persistence length 50nm.  If we neglect the excluded volume, electrostatic, and solvent-induced interactions of the DNA (a rather severe approximation), and take $\r\approx 0.9$ (the saturating value of $\r$, see Fig. \ref{rhoplotFig}a), we find $P\approx 1$kPa (=$10^{-3}$pN/nm$^2$), almost 4 orders of magnitude lower than the 6MPa measured in the experiments.  It is clear that the behavior of a strongly confined wormlike chain is critically dependent on the intra-chain interactions, in agreement with a number of other studies \cite{VirusEnergy1,VirusEnergy2,DiscreteVirus,TorroidVirus1,MuthVirus,BloomfieldEnergy}.  
While we have found that entropy of confinement produces a negligible contribution to the experimentally observed pressure, excluded volume interactions will further restrict the conformational space available to the chain.  Intra-chain interactions will lead to an increase in the entropic contribution to the free energy and pressure, as has been seen in simulations \cite{HarveyPhi29}.  Our results establish firmly, as noted some time ago \cite{BloomfieldEnergy}, that the origin of spool-like order and the extremely large pressure of DNA in a capsid, is due to inter-segment and counterion-mediated interactions.

\section{Conclusions}

We have shown that WLCs in restricted spaces can be accurately treated by applying the mean field theory \cite{MeanField1,MeanFieldForce,THBook}.  For a surface confined chain we can determine many average properties of the WLC, by replacing the rigid constraints of inextensibility ($\Dr_n^2\equiv a^2$) and confinement ($\rv_n^2\equiv R^2$) with average constraints.  We have shown that the mean field approach reproduces the exact results of Spakowitz and Wang \cite{SWSphereSurf}, and reproduces the correct scaling coefficient of the free energy of confinement.  The mean field approach is also able to 
determine the scaling and free energy of a surface confined WLC under tension, which may be of use in better understanding the wrapping of DNA around histones \cite{Histone1,Histone2}.  The force-extension curve (FEC) for a strongly confined WLC differs greatly from the unconfined FEC, with oscillatory behavior.  

We also find that the mean field method can approximately determine the behavior of a WLC confined to the interior of a sphere.  Interior monomers (far from the endpoints) are approximately surface confined, with $\la\rv^2(s)\ra\approx 0.9 R^2$ for strongly confined chains, but endpoint effects dominate the behavior of the chain for $s<2R$ or $s>L-2R$.  Structural information about the confined chain can be determined by examining the correlations in the local winding axis, and we find that strongly confined stiff chains are highly structured, even without intra-chain interactions.  The mean field estimates of the pressure due to confinement show that the extreme pressures inside of a viral capsid are not strongly dependent on simple confinement entropy, but must arise from intra-chain and counterion-mediated interactions.  The good agreement with the simulated pressures allows us to accurately estimate the free energy of confinement of a strongly confined WLC arising from energetic considerations alone as
\begin{eqnarray}
\beta F\approx 0.56\frac{Ll_p}{R^2}-1.1\frac{l_p}{R}+3\log(R)
\end{eqnarray}
The excellent agreement between theory and simulations show that the mean field theory can be adapted to include thee effects of inter-segment interactions, even when semiflexible chains are confined to restricted spaces.

\setcounter{section}{0}
\def\thesection       {\Alph{section}}

{\section{The $\Q$ Matrix}}
\label{QMatApp}

Defining $\xv^N=(x_1,\dots,x_{N+1})$, we can rewrite the Hamiltonian in Eq. \ref{onedconf} as
\begin{eqnarray}
{\H_x}=a\sum_n\blp\frac{l_p}{2}\ \frac{(\Delta x_{n+1}-\Delta x_n)^2}{a^4}+\l_n\frac{\Delta x_n^2}{a^2}+k_n \frac{x_n^2}{R^2}\brp =\left(\xv^N \right)^T\Q\xv^N
\end{eqnarray}
where the elements of the tridiagonal matrix $\Q$ are:
\begin{eqnarray}
\Q_{i,i+2}=\Q_{i+2,i}&=&\frac{l_p}{2a^4}\label{qeq1}\\
\Q_{i,i+1}=\Q_{i+1,i}&=&\left\{\begin{array}{cc}-\frac{l_p}{a^4}-\frac{2\l_i}{a^2} & i=1,N \\ -\frac{2l_p}{a^4}-\frac{2\l_i}{a^2} & \mbox{else}\end{array}\right.\\
\Q_{i,i}&=&\left\{\begin{array}{cc}\frac{l_p}{2a^4}+\frac{\l_1}{a^2} +\frac{a k_1}{R^2}& i=1 \\\frac{l_p}{2a^4}+\frac{\l_N}{a^2} +\frac{a k_{N+1}}{R^2}& i=N+1 \\\frac{5l_p}{2a^4}+\frac{\l_{i-1}+\l_i}{a^2}+\frac{ak_i}{R^2} & i=2,N \\\frac{3l_p}{a^4}+\frac{\l_{i-1}+\l_i}{a^2}+\frac{ak_i}{R^2} & \mbox{else}\end{array}\right.\label{qeq3}
\end{eqnarray}
The structure of $\Q$ is unchanged under the transformation $k_1=k_{N+1}$, $k_2=k_N$, $\l_1=\l_N$, $k_i=k$ for $2<i<N$ and $\l_i=\l$ for $1<i<N$.

{\section{Evaluation of the 1-D Confined Propagator}}
\label{MMatApp}
We are interested in evaluating the path integral in Eq. \ref{Fx}, 
\begin{equation}
Z_0(\xv)=\int\D[x(s)]\exp\blp-\frac{\lt}{2}\int_0^{\Lt} d\s\ \ddot{x}^2(s)-\L\int_0^{\Lt} ds\  \dot x^2(s)-\frac{k}{R^2}\int_0^{\Lt} ds\ x^2(s)\brp,
\end{equation}
subject to the boundary conditions $x(0)=x_0$, $u(0)=u_0$, $x(L)=x_L$, and $u(L)=u_L$.  
We write $x(s)=f(s)+g(s)$, where $g(0)=g(\Lt)=\dot g(0)=\dot g(\Lt)=0$ and where
\begin{eqnarray}
\frac{\lt }{ 2}f^{(4)}(s)-\L \ddot f(s)+\k f(s)=0\nonumber,
\end{eqnarray}
with $f(0)=x_0,\ f(\Lt)=x_L,\ \dot f(0)=u_0,\ \mbox{and }f(\Lt)=u_L$.  If $f$ satisfies the above differential equation, a simple integration by parts gives
\begin{eqnarray}
Z_0(\xv)&=&K(\Lt)\exp\blp -\frac{\lt}{2}\bigg[\ddot f\dot f-f^{(3)}f\bigg]_0^{\Lt}-\L[u_L x_L-u_0x_0]\brp\label{Zoeq}\\
K(\Lt)&=&\int\D[g]\exp\blp-\frac{\lt }{ 2}\int_0^{\Lt} ds\ \ddot{g}^2(s)-\L\int_0^{\Lt} ds\  \dot g^2(s)
-\frac{\k}{R^2}\int_0^{\Lt} ds\ g^2(s)\brp\label{KdefInt}
\end{eqnarray}
where $g$ and $\dot g$ vanish at the boundaries.  
The exponential term in Eq. \ref{Zoeq} can be evaluated by solving the differential equation for $f$ directly, giving 
\begin{equation}
Z_0(\xv)=K(\Lt)e^{-\xv\cdot\M\xv},
\end{equation}
where
\begin{eqnarray}
\M=\frac{\lt R}{ 2 d(\Lt)}\left(
\begin{array}{cccc}
R\,m_{11} & R\,m_{12} & R\,m_{13} & m_{14} \\ R\,m_{12} & R\,m_{11} & -m_{14} & -R\,m_{13} \\ R\,m_{13} & -m_{14} & m_{33}/R & m_{34}/R \\ m_{14} & -R\,m_{13} & m_{34}/R & m_{33}/R
\end{array}
\right) 
+\frac{\L}{2}\left(
\begin{array}{cccc}
0 & 0 & -1 & 0 \\ 0 & 0 & 0 & 1 \\ -1 & 0 & 0 & 0 \\ 0 & 1 & 0 & 0
\end{array}
\right) . 
\label{Meq}
\end{eqnarray}
In Eq. \ref{Meq}, we have defined
\begin{eqnarray}
d(\Lt)&=&\frac{2\w_1 \w_2(1-\co\ct)+(\w_1^2+\w_2^2)\so\st }{\wo\wt(\w_1^2-\w_2^2)}\nonumber\\
m_{11}&=&\wo\so\ct-\wt\co\st\nonumber\\
m_{12}&=&\wt\st-\wo\so\nonumber\\
m_{13}&=&\frac{\wo^2-\wt^2 }{ 2 \wo \wt}\so\st\nonumber\\
m_{14}&=&\co-\ct\nonumber\\
m_{33}&=&\frac{1 }{ \wt}\co\st-\frac{1 }{ \wo}\so\ct\nonumber\\
m_{34}&=&\frac{1 }{ \wo}\so-\frac{1 }{ \wt}\st\nonumber,
\end{eqnarray}
with
\begin{equation}
\w_i=\left[\frac{\L }{ \lt}\blp 1\pm \sqrt{1-\frac{2\k\lt }{ \L^2}}\brp\right]^{\frac{1}{2}}.
\end{equation}

Note that the full propagator $Z(\xv)$ in Eq. \ref{Zdef} can be written as $Z(\xv)=Z_0(\xv)\,\exp(-\xv\cdot\G\xv)$, with the matrix $\G$ containing terms suppressing excess endpoint fluctuations,
\begin{eqnarray}
\G= \left(
\begin{array}{cccc}
\g_1/R^2 & 0 & \g_2/R & 0 \\ 0 & \g_1/R^2 & 0 & -\g_2/R \\ \g_2/R & 0 & \d & 0 \\ 0 & -\g_2/R & 0 & \d
\end{array}
\right) 
\end{eqnarray}

In general, computing average values involves calculating the determinant of $\M+\G$.  Simplification of the determinant is a tedious process, but it is useful to note that 
\begin{eqnarray}
\Det(\M)&=&A_1^2-A_2^2\\
A_1&=&m_{13}^2+m_{14}^2+m_{12}m_{34}-m_{11}m_{13}\nonumber\\
A_2&=&2m_{13}m_{14}+m_{12}m_{33}-m_{11}m_{34}\nonumber
\end{eqnarray}
with a similar relation holding for Det$(\M+\G)$.

We can calculate $K(\Lt)$ by the evaluation of a simple integral.  Following the standard method of Feynman \cite{Feyn}, we can write the propagator from $(x_0,u_0)$ to $(x_L,u_L)$ as an integral over all intermediate points, $(x_s,u_s)$,
\begin{eqnarray}
Z_0(x_0,x_L,u_0,u_L;L)&=&\int_{-\infty}^\infty dx_sdu_s\,Z_0(x_0,u_0,x_s,u_s;s)\times Z_0(x_s,u_s,x_L,u_L;L-s)\nonumber\\
&=&\int_{-\infty}^{\infty}\int_{-\infty}^\infty dx_s du_s\ K(s)\exp\blp-\xv_1^T\cdot\M(s)\xv_1\brp\label{FindK}\\
&&\qquad\qquad\qquad\times K(L-s)\exp\blp-\xv_2^T\cdot\M(L-s)\,\xv_2\brp\nonumber
\end{eqnarray}
where $\xv_1=(x_0,x_s,u_0,u_s)$ and $\xv_2=(x_s,x_L,u_s,u_L)$.  $\M$ has already been determined (Eq. \ref{Meq}), and $K(L)$, given in Eq. \ref{KdefInt}, is independent of all $x_i$'s and $u_i$'s.  The integral in Eq. \ref{FindK} is tedious to evaluate, but yields
\begin{eqnarray}
Z_0(\xv;L)&\equiv& K(\Lt)e^{-\xv^T\cdot\M(L)\,\xv} \nonumber\\
&=&K(s)K(\Lt-s)\ \frac{2\pi}{\lt} \blp \frac{d(s)d(\Lt-s) }{(\wo^2-\wt^2)d(\Lt) }\brp^{\frac{1 }{ 2}}e^{-\xv^T\cdot\M(L)\,\xv}.
\end{eqnarray}
We then find 
\begin{equation}
K(\Lt)=\frac{\lt }{ 2\pi}\sqrt{\frac{\wo^2-\wt^2 }{ { d(\Lt)}}}\ e^{-\Lt \eta},\label{Keq}
\end{equation}
where $\eta$ is an arbitrary constant.


In the limit of strong confinement ($L\w_i\gg1$, see the main text),
\begin{eqnarray}
\M&=&
\frac{l_pR\w_1\w_2}{2}\left(
\begin{array}{cccc}
R(\w_1+\w_2) & 0 & 1 & 0 \\ 0 & R(\w_1+\w_2) & 0 & -1 \\ 1 & 0 & (\w_1^{-1}+\w_2^{-1})/R & 0 \\ 0 & -1 & 0 & (\w_1^{-1}+\w_2^{-1})/R
\end{array}
\right) \nonumber\\
K&\propto&\sqrt{\w_1\w_2}\,(\w_1+\w_2)e^{-L(\w_1+\w_2)/2}
\end{eqnarray}
This strongly confined representation is significantly easier to work with when computing the mean field solutions.

To ensure that our calculation of $Z_0$ has the correct limiting behavior, we find 
\begin{eqnarray}
\lim_{R\to\infty}\int_{-\infty}^\infty dx_L Z_0(\xv)=e^{-kLx_0^2}\ \blp\frac{l_p\O}{2\pi\sinh(\O L)}\brp^{\frac{1}{2}}\exp\blp-\frac{l_p\O}{2\sinh(\O L)}\qquad\qquad\\
\times\bigg[(u_0^2+u_L^2)\cosh(\O L)-2u_0u_L\bigg]+O(1/R)\brp\nonumber,
\end{eqnarray}
with $\Omega=\sqrt{2\l/l_p}$,  identical to the unconfined propagator found in the work of Ha and Thirumalai \cite{MeanField1,THBook}, except for the the term $e^{-k L x_0^2}$.  Since $k\propto R^{-2}$ in all cases considered, the integration over $x_0$ leads to a divergent integral.  However,  as $R\to\infty$, the system becomes translationally invariant, so that integration over the initial position will be proportional to the radius of the confinement.  
%
The integral over the initial position then simply adds an irrelevant constant to the free energy, and we can write the one dimensional propagator
\begin{eqnarray}
Z_U(u_0,u_L)=\lim_{R\to\infty}\frac{1}{R}\int_{-\infty}^\infty dx_0 dx_L \ Z(\xv)
\end{eqnarray}
identical to the result found by Ha and Thirumalai \cite{MeanField1,THBook} up to a multiplicative constant.

\section{Details of the simulations for volume confinement}
\label{SphereSimApp}

We have therefore performed a number of Langevin Dynamics simulations with varying $L$, $l_p$, and $R$ when considering volume confinement.  The Hamiltonian used is
\begin{eqnarray}
\beta H=\frac{k}{2a^2}\sum_{i=0}^{N}(|\rv_{i+1}-\rv_i|-a)^2-\frac{l_p}{a}\sum_{i=0}^{N-1}\uv_i\cdot\uv_{i+1}+\e_S\sum_{i=0}^{N+1}\blp\frac{a}{|\rv_i|-(R+a)}\brp^{12}\label{simHam}
\end{eqnarray}
with $L=Na$.  The first term ensures the connectivity of the chain, and we take $k=10^4$ throughout, ensuring very stiff bonds.  The second term accounts for the bending stiffness of the chain, with persistence length $l_p$.  We have confirmed directly that this Hamiltonian in the unconfined case (i.e. only the first two terms of Eq. \ref{simHam} are used) gives $\la\uv(s)\cdot\uv(s')\ra=e^{-|\Delta s|/l_p}$ to within $\sim 5\%$.   The third term of the Hamiltonian approximately confines the chain to the interior of a sphere of radius $R$, using a Lennard-Jones repulsion.  The confinement energy is on the order of $\e_S\,k_BT$ when $|\rv_i|=R$, and increases sharply for larger $|\rv_i|$.  We choose $\e_S=1$ throughout the simulations, which restricts $|\rv_i|/R\lesssim 1.01$ for all of the parameters we considered.  To determine the equilibrium properties of the system, we use the low friction limit \cite{HoneycuttBP92}, with $\eta = 0.1$, and a timestep of $h=0.001$ (in dimensionless units, or equivalently with the mass $m=1$, spacing $a=1$, and $k_B T=1$).  In the simulations, we consider a chain with $N=200$ for $l_p/a=20$, 50, and 100, with$R/a=5$, 6, 7, and 8.  We also consider a chain with $N=100$ for $l_p/a=5$, 10, 20, 50, and 100, with $R/a=5$, 6, 7, 8, 9, 10, 11, 12, 15, 20, and 25.


\section{Calculation of the winding axis}
\label{WindingApp}
Because of the spherical symmetry of the problem, we are free to choose our coordinate system such that it simplifies the calculation.  We take $\hat\uv_{i-1}=\hat {{\mathbf{z}}}$, defining the $z$-axis, and $\hat\uv_i=(\sin[\Theta_i],0,\cos[\Theta_i])$, defining the $x$-axis.  We take our third bond to be $\hat\uv_{i+1}=(\sin[\Theta_{i+1}]\cos[\j],\sin[\Theta_{i+1}]\sin[\j],\cos[\Theta_{i+1}])$.  With $\theta_{i,j}$ the angle between bonds $i$ and $j$, we see $\theta_{i-1,i}=\Theta_i$ and $\theta_{i-1,i+1}=\Theta_{i+1}$.  It is convenient to eliminate the azimuthal angle $\j$ when computing $\cos(\theta_{i,i+1})=\hat\uv_i\cdot\hat\uv_{i+1}$, giving
\begin{eqnarray}
\cos(\j)=\cot(\theta_{i,i+1})\cot(\Theta_i)-\cos(\Theta_{i+1})\csc(\theta_{i,i+1})\csc(\Theta_i)
\end{eqnarray}
In this coordinate system, $\hat\av_i=\hat{\mathbf{y}}$, and $|\av_{i+1}|=\sin(\theta_{i,i+1})$.    To compute the dot product between the two winding axes, we need only $\av_{i+1}\cdot\hat{\mathbf{y}}=\cos(\Theta_i)\sin(\Theta_{i+1})\cos(\j)-\cos(\Theta_{i+1})\sin(\Theta_i)$.  Eq. \ref{WindingAxisAngles} is recovered upon substitution of $\cos(\j)$ in $\hat\av_i\cdot\hat\av_{i+1}=\av_{i+1}\cdot\hat{\mathbf{y}}/|\av_{i+1}|$.\\

{\bf{Acknowledgments:}}  This work was supported in part by a grant from the National Science Foundation through NSF CHE 05-14056.

%


\clearpage

{\bf{Figure Captions}}

{\underline{Fig 1}}:  Representative structures for a WLC confined to the surface of a sphere of radius $R=3a$.  (a) shows $l_p=2.5a$ and (b) shows $l_p=20a$.  An enlargement of the polymer in (b) diagrams the positions $\rv_i$ and bond vectors $\uv_i$.

{\underline{Fig. 2}}:  Free energy $\beta\Delta F=\beta F(R)-\beta F(\infty)$ as a function of $R$ for a surface-confined WLC.  $\beta F(\infty)$ is determined from a simulation with $R=2\times 10^4a$.  The symbols are the simulation data, where the lines are the theoretical results of eq. \ref{surfF}.  Shown are $l_p/a$=20 (solid purple), 10 (dotted blue), 5 (dashed green), and 2.5 (dot-dashed red).  The inset shows $\log(\beta \Delta F)$ as a function of $R$, displaying the good agreement between simulation and theory, particularly for large $R$.

{\underline{Fig. 3}}:  Linear extension under an external tension of a surface confined WLC as a function of the radius.  In (a), $L=450a$ and $l_p=150a$.  The applied tensions are $a\beta f$= 0.1 (solid blue), 0.05 (dashed green) and 0.01 (dotted red), displaying the oscillations in $\la x_L-x_0\ra$ for stiff chains.  In (b), the same values of the force are applied to a chain of length $L=450a$ and $l_p=15a$, showing that the FEC of a flexible chain is monotonic.

{\underline{Fig. 4}}:  $\la\rv^2(s)\ra$ vs. $s$.  (a):  $L=100a$ and $l_p=100a$, with (from highest to lowest) $R/a$=5, 10, and 20.   The average monomer position is dominated by endpoint effects for $s\lesssim2R$ and $s\gtrsim L-2R$.  (b):  $L=200a$ and $R=5a$, with (from highest to lowest) $l_p/a$=100, 50, and 20.  With $R$ fixed, variations in $l_p$ change only the value of $\rho$, but do not alter the behavior of the monomers. 

{\underline{Fig. 5}}:  (a)  The mean field parameters $\r$ (solid blue), $\r_0$ (dotted red), and $\r_c$ (dashed green) as a function of $l_p/R$.  Lines are determined from a simulations with $L=100a$ for various $l_p$ and $R$.  Points are from simulations with $L=200a$.  (b)  $\alpha=l_p/l_0$ as a function of $l_p/R$, determined by fitting eq. \ref{bend_volume} to the simulation results.  Symbols are the fits for $L=100a$, the line is the fit for $L=200a$.

{\underline{Fig. 6}}:  Bending correlation function $\la\uv(L/2)\cdot\uv(s)\ra$ as a function of $s$ for a chain with $L=200a$.  The points are simulation data, the solid lines are the theoretical results in Eq. \ref{bend_volume}.  (a) has $l_p=20a$ and $R=8a$, with $\a\approx 2$, and (b) has $l_p=100a$ and $R=5$ with $\alpha \approx 5/2$.  The agreement between theory and simulations is excellent, except near the endpoints. Representative configurations are shown to the right, and clearly displays the wrapping to the chain for a strongly confined WLC in (b).

{\underline{Fig. 7}}:  (a) Average correlations in the nearest neighbor winding axis as a function of $al_p/R^2$, with the average taken over interior points only (i.e. $2R\le s\le L-2R$).  The symbols are simulation data for $L=100a$ for various $l_p$ and $R$, with the line the simulation data for $L=200a$.  The inset shows the average nearest neighbor correlation for $L=200a$, $R=5a$, and $l_p/a=100$ (blue), 50 (green), and 20 (red).  The correlations drop sharply near the endpoints.  (b)  $\la\hat\av(L/2)\cdot\hat\av(s)\ra$ as a function of $s$ for $L=200a$ and $R=5a$, for $l_p/a$=100, 50, and 20.  The curves are a fit to the exponential$\la\hat\av(L/2)\cdot\hat\av(s)\ra\propto e^{-L/2l_0}$.  One representative configuration for $L=200a$ and $l_p=100a$, seen from two different viewpoints, clearly shows high correlations in the winding axis, except near the ends of the chain.

{\underline{Fig. 8}}:  $\beta P V$ as a function of $R$ for varying $l_p$ and $L$.  In both, the dots are simulation results, and the solid line is the theoretical result, with $\rho$, $\r_0$ and $\r_c$ taken directly from the simulation results. (a):  $L=100 a$ and $l_p/a$=100 (red), 50 (green), 20 (blue), 10 (purple), and 5 (pink).  The inset is a log-log plot for this data.  (b):  $L=200a$ and $l_p/a$=100 (red), 50 (green), and 20 (blue).

\clearpage

\renewcommand{\baselinestretch}{2}
\small\normalsize

\clearpage

\begin{figure}
\centering
\includegraphics[width=0.7\textwidth]{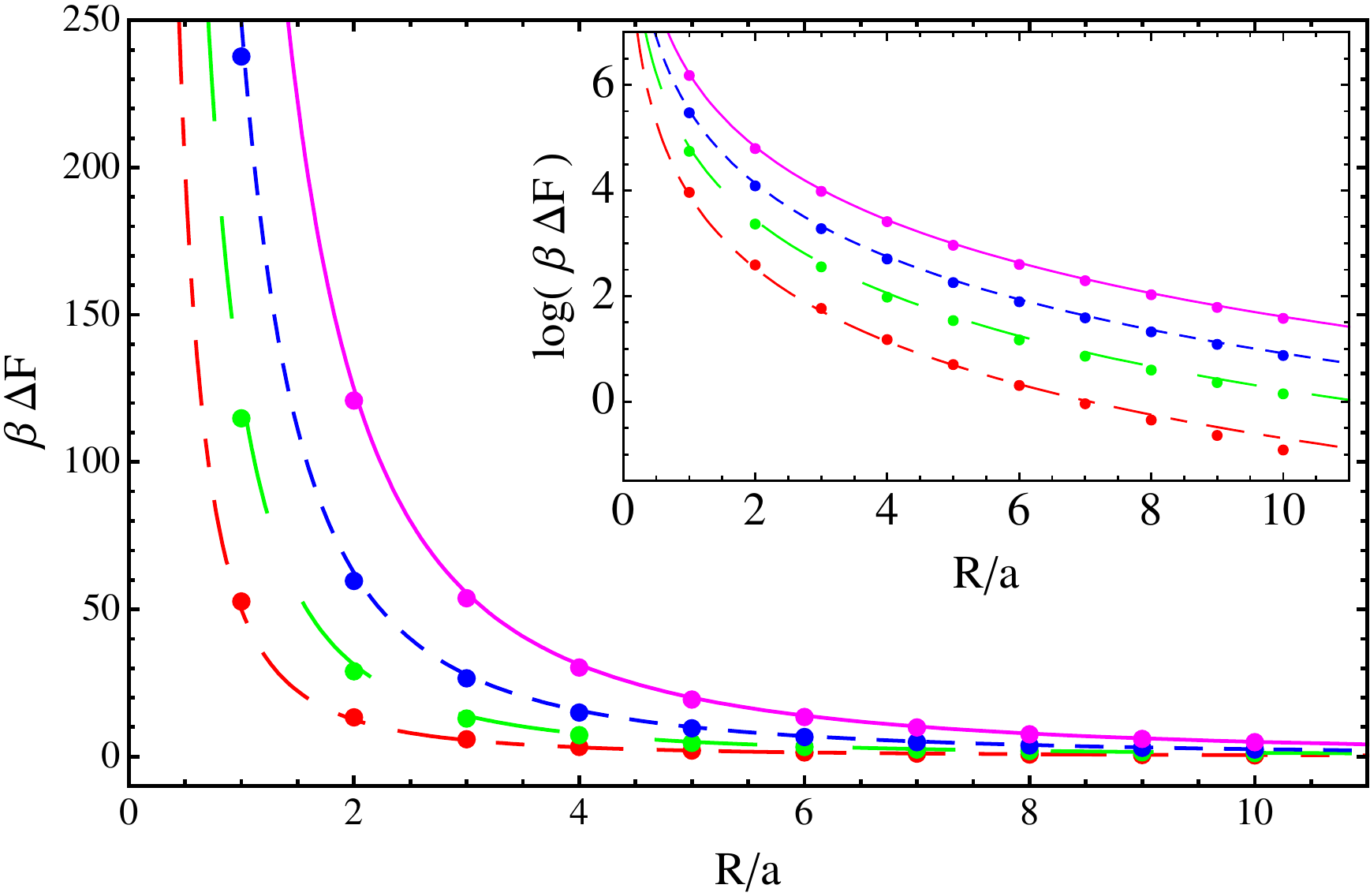}
\renewcommand{\baselinestretch}{1}
\small\normalsize
\caption{}
\label{SurfF}
\end{figure}
\renewcommand{\baselinestretch}{2}
\small\normalsize

\clearpage

\begin{figure}[htbp]
\begin{center}
\includegraphics[width=.9\textwidth]{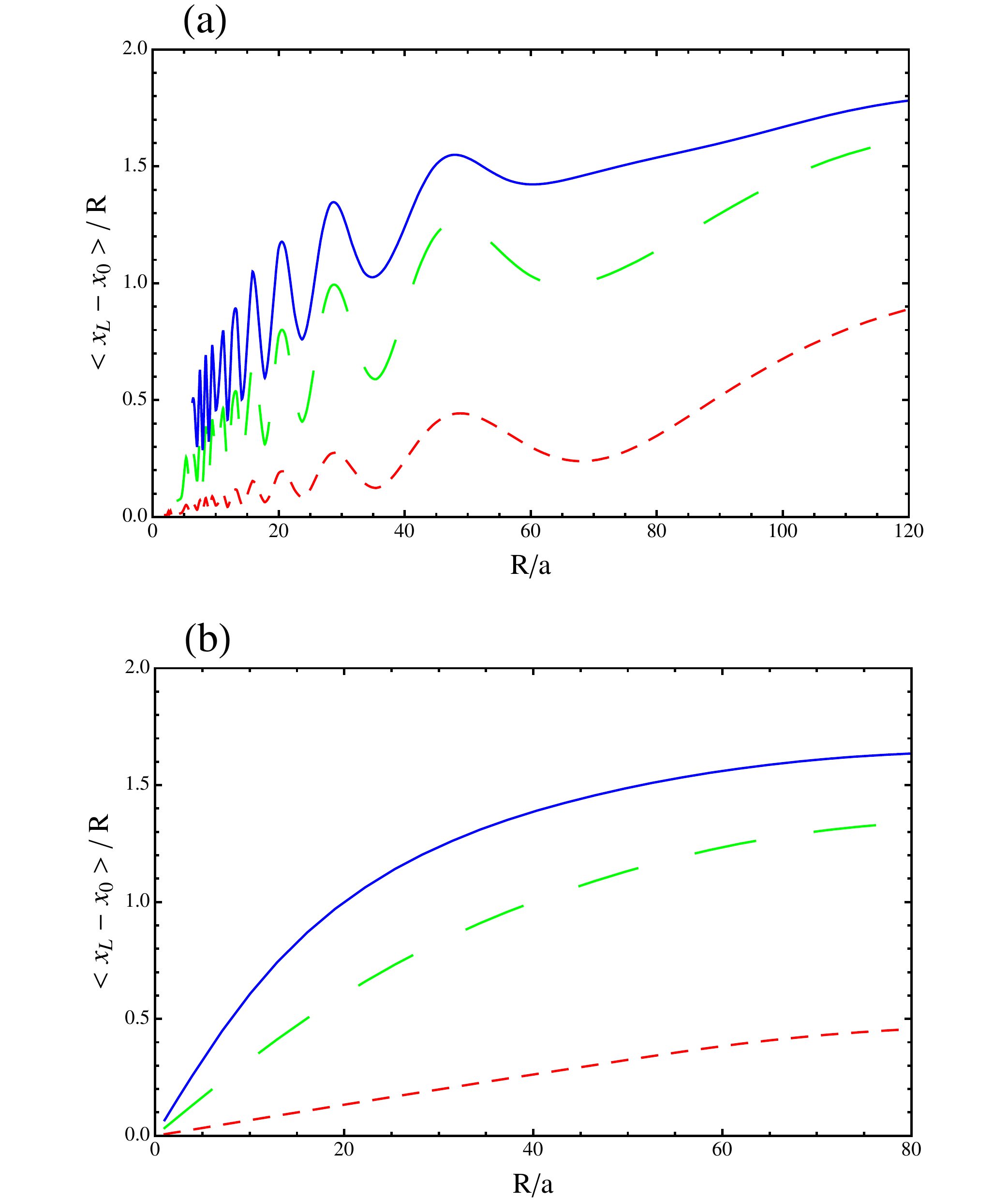}
\renewcommand{\baselinestretch}{1}
\small\normalsize
\caption{}
\label{FEC_SurfFig}
\end{center}
\end{figure}
\renewcommand{\baselinestretch}{2}
\small\normalsize

\clearpage

\begin{figure}[tbp]
\begin{center}
\includegraphics[width=.9\textwidth]{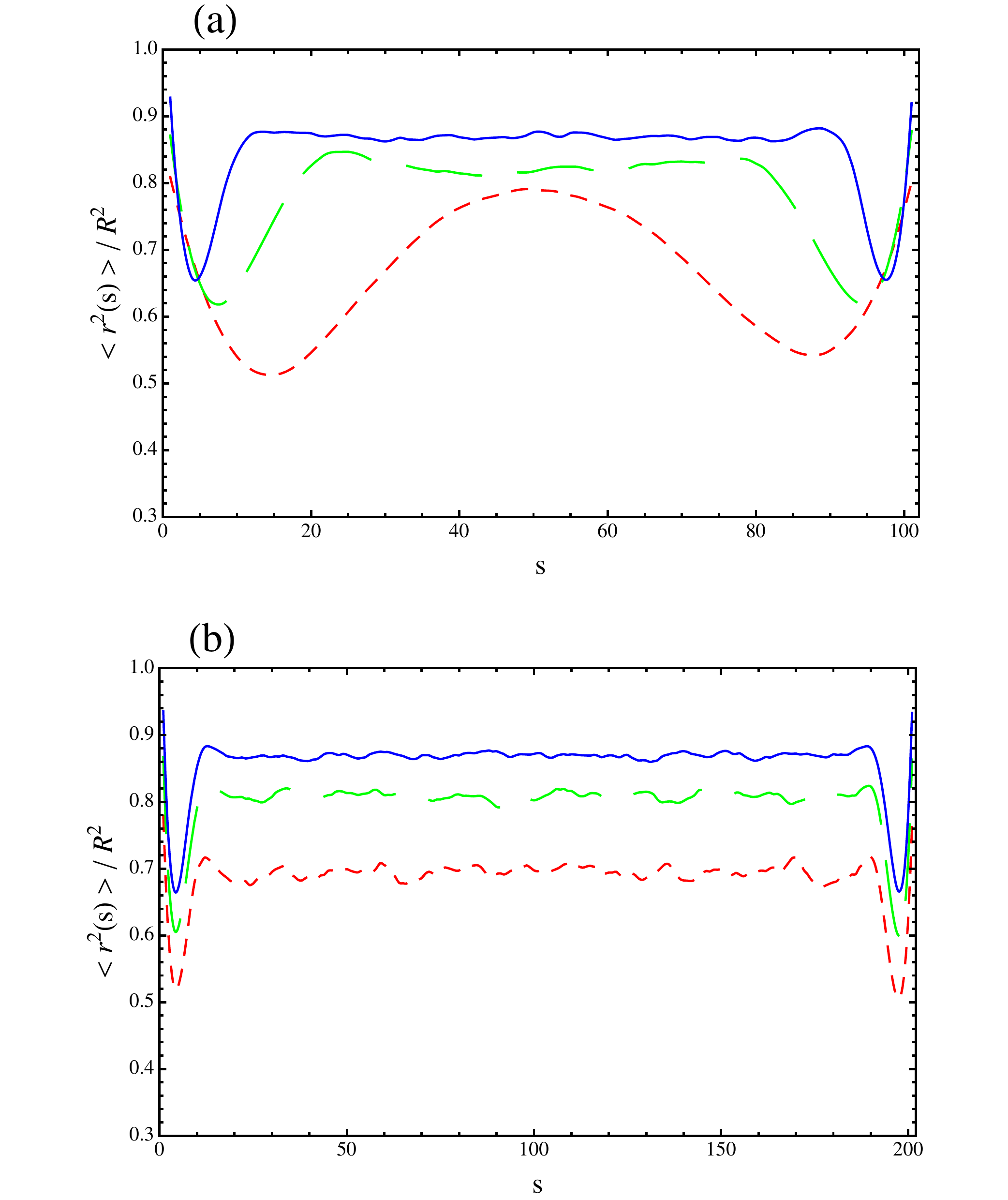}
\renewcommand{\baselinestretch}{1}
\small\normalsize
\caption{}
\label{rmagFig}
\end{center}
\end{figure}
\renewcommand{\baselinestretch}{2}
\small\normalsize

\clearpage

\begin{figure}[tbp]
\begin{center}
\includegraphics[width=.9\textwidth]{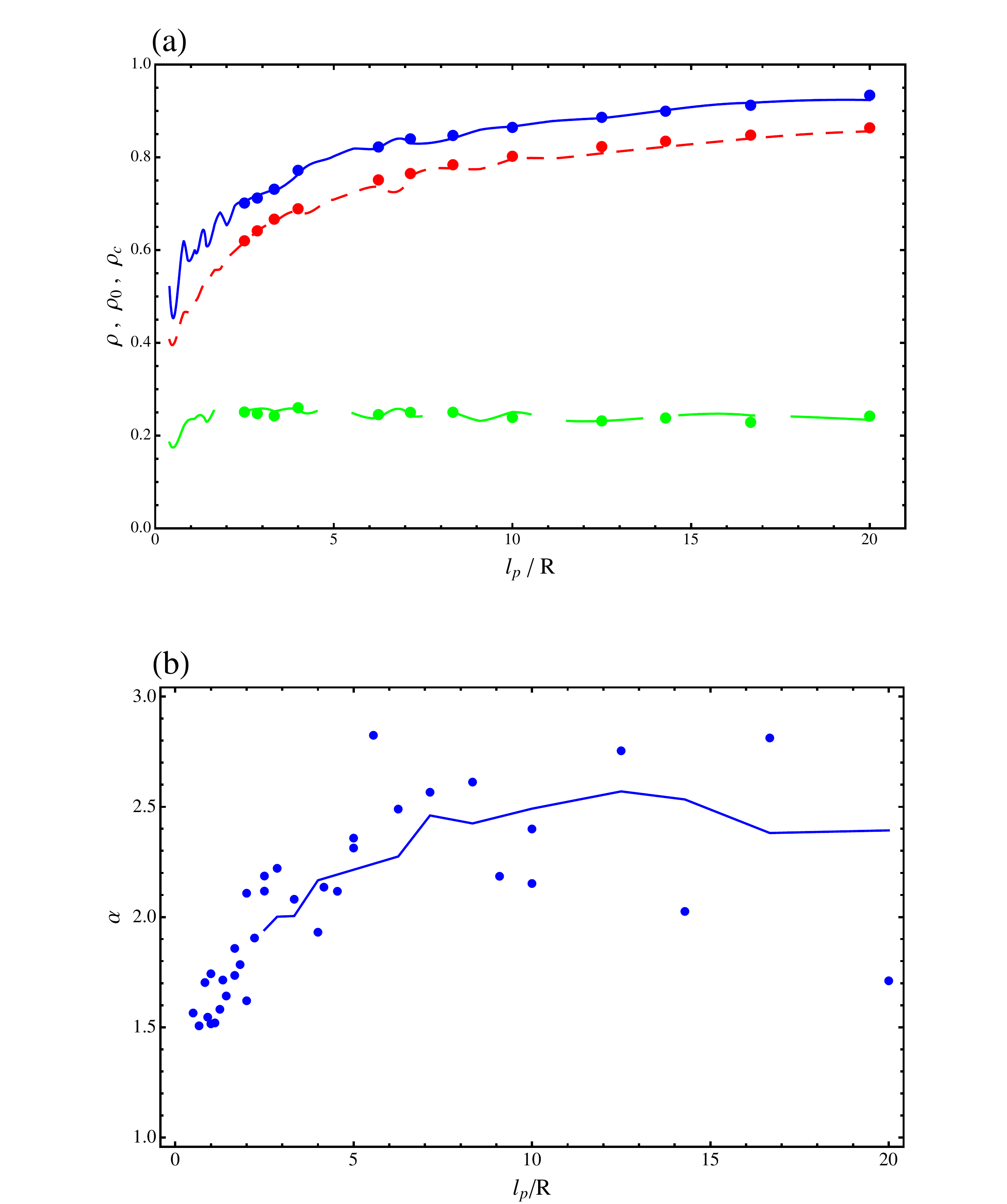}
\renewcommand{\baselinestretch}{1}
\small\normalsize
\caption{} 
\label{rhoplotFig}
\end{center}
\end{figure}
\renewcommand{\baselinestretch}{2}
\small\normalsize

\clearpage

\begin{figure}[htbp]
\begin{center}
\includegraphics[width=.9\textwidth]{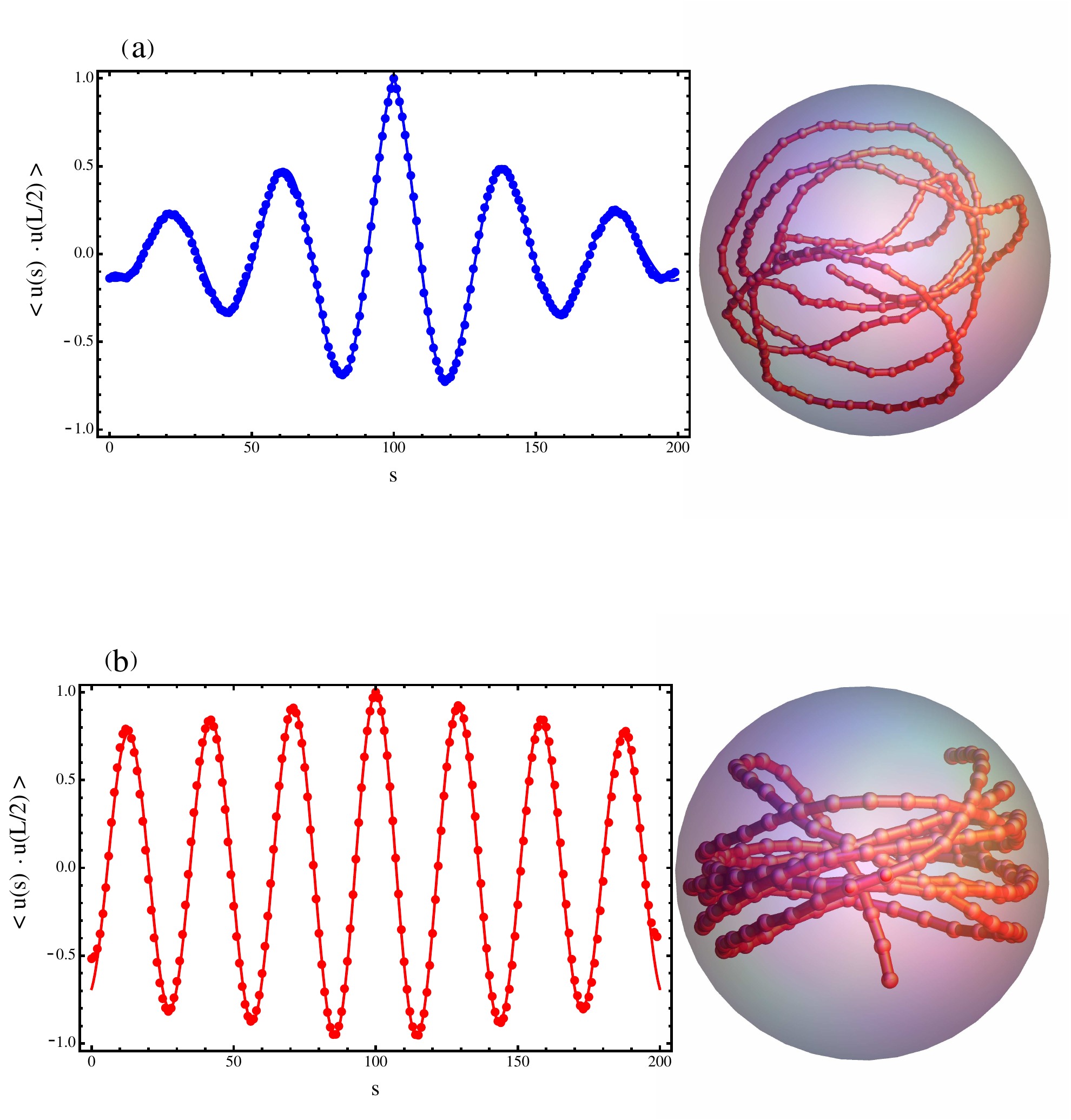}
\renewcommand{\baselinestretch}{1}
\small\normalsize
\caption{}  

\label{bendCorrFig}
\end{center}
\end{figure}
\renewcommand{\baselinestretch}{2}
\small\normalsize

\clearpage

\begin{figure}[htbp]
\begin{center}
\includegraphics[width=.6\textwidth]{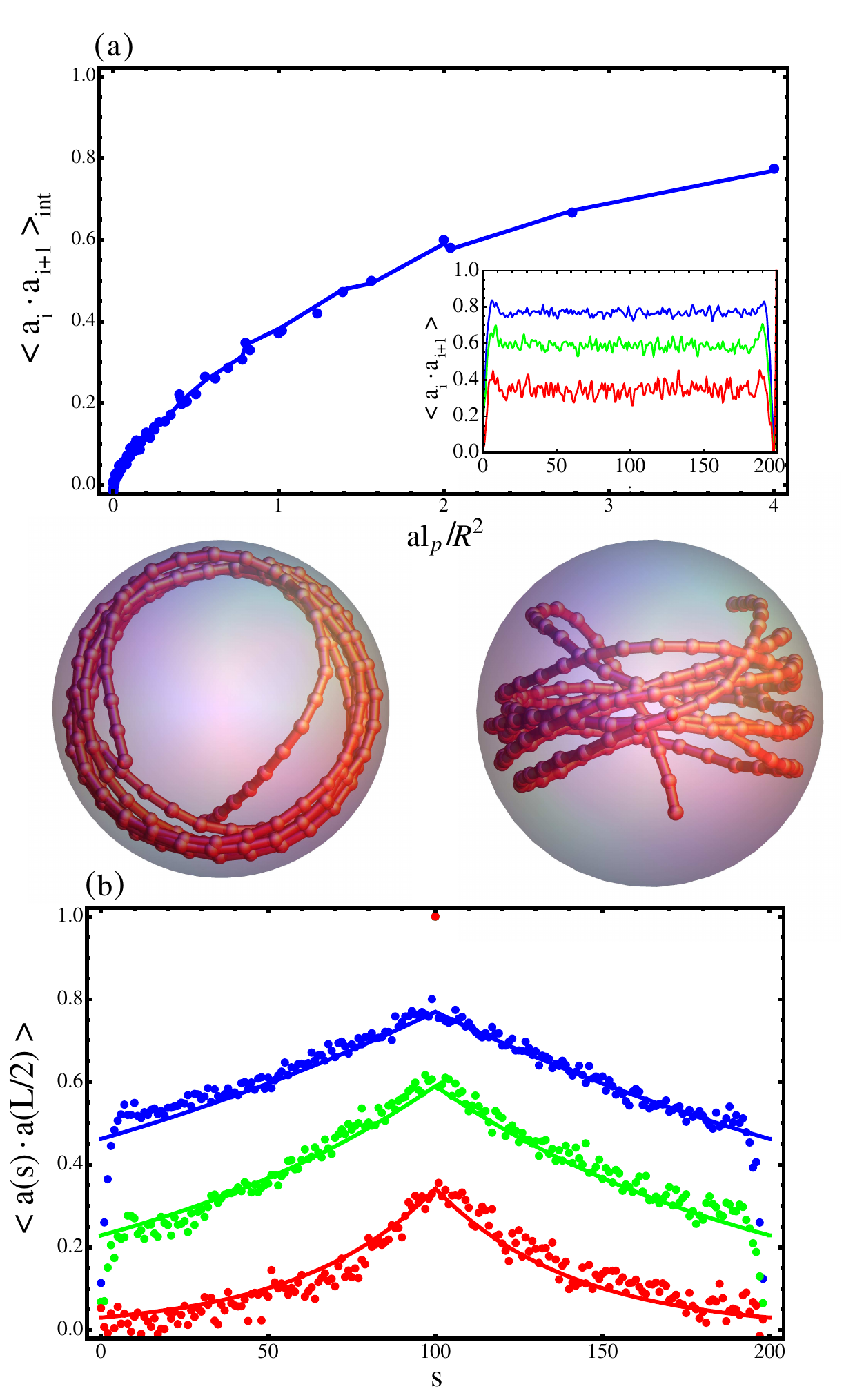}
\renewcommand{\baselinestretch}{1}
\small\normalsize
\caption{}
\label{windFig}
\end{center}
\end{figure}
\renewcommand{\baselinestretch}{2}
\small\normalsize

\clearpage

\begin{figure}[htbp]
\begin{center}
\includegraphics[width=.9\textwidth]{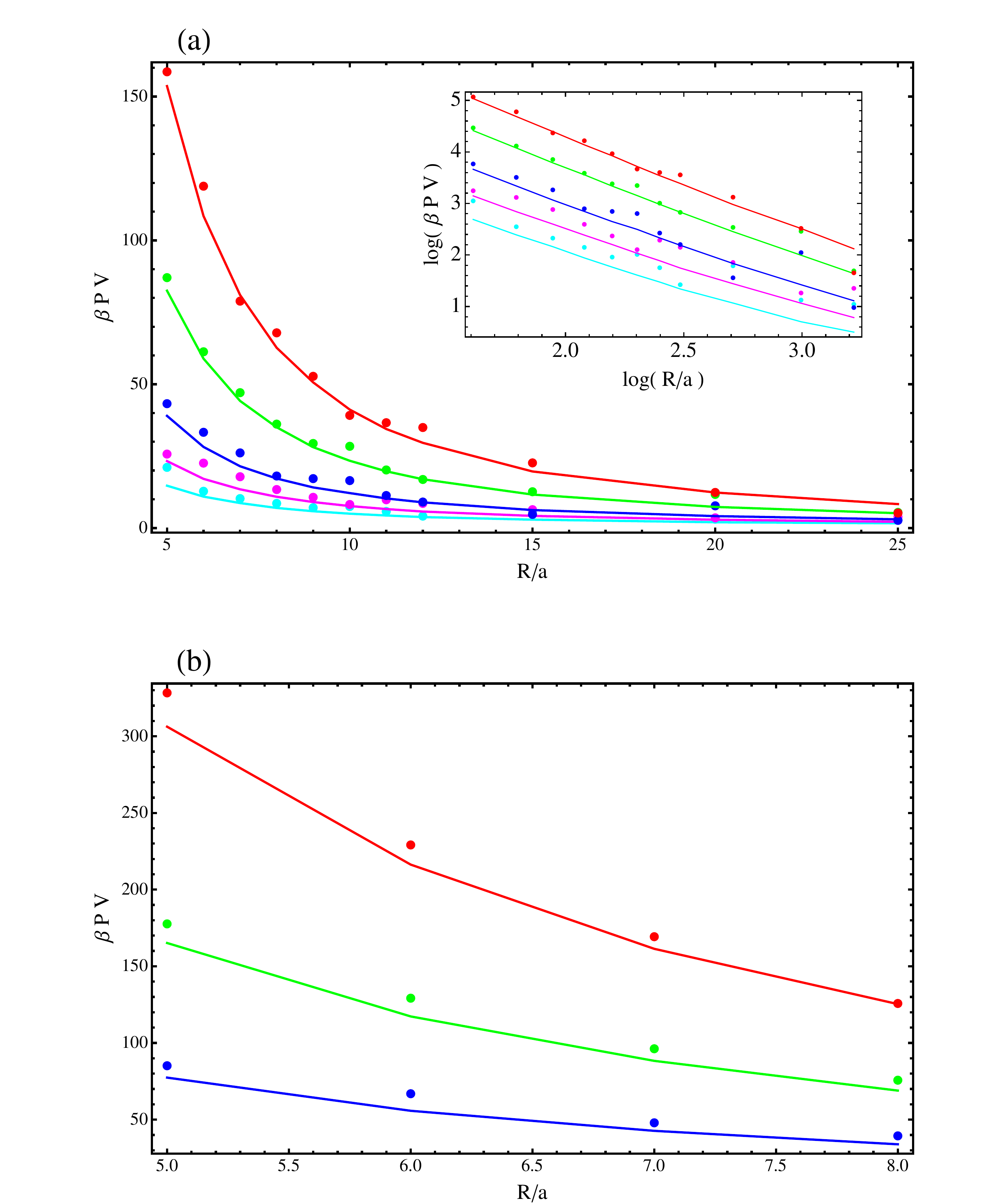}
\renewcommand{\baselinestretch}{1}
\small\normalsize
\caption{}
\label{pressFig}
\end{center}
\end{figure}
\renewcommand{\baselinestretch}{2}
\small\normalsize

\clearpage

\renewcommand{\baselinestretch}{1}


\small\normalsize		



\begin{thebibliography}{10}

\bibitem{KP}
O.~Kratky and G.~Porod.
\newblock {\em Rec. Trav. Chim.}, 68:1106, 1949.

\bibitem{Histone1}
I.~M. Kuli\'c and H.~Schiessel.
\newblock {DNA} spools under tension.
\newblock {\em Phys. Rev. Lett.}, 92:228101, 2004.

\bibitem{Histone2}
S.~Mihardja, A.~J. Spakowitz, Y.~Zhang, and C.~Bustamante.
\newblock Effect of force on mononucleosomal dynamics.
\newblock {\em Proc. Natl. Acad. Sci.}, 103:15871, 2006.

\bibitem{HistoneManning}
N.~Marky and G.~S. Manning.
\newblock A theory of {DNA} dissociation from the nucleosome.
\newblock {\em J. Mol. Biol.}, 254:50--61, 1995.

\bibitem{OdijkSurface}
T.~Odijk.
\newblock Physics of tightly curved semiflexible polymer chains.
\newblock {\em Macromolecules}, 26:6897, 1993.

\bibitem{SurfaceEnergyMin}
C.~Lin, Y.~Tsai, and C.~Hu.
\newblock Wrapping conformations of a polymer on a curved surface.
\newblock {\em Phys. Rev. E}, 75:031903, 2007.

\bibitem{FJCSurface}
R.~Mondescu and M.~Muthukumar.
\newblock Brownian motion and polymer statistics on certain curved manifolds.
\newblock {\em Phys. Rev. E}, 57:4411, 1998.

\bibitem{SWSphereSurf}
A.~Spakowitz and Z.~Wang.
\newblock Semiflexible polymer confined to a spherical surface.
\newblock {\em Phys. Rev. Lett.}, 91:166102, 2003.

\bibitem{SWSphereSim}
J.~Cerd\'a, T.~Sintes, and A.~Chakrabarti.
\newblock Excluded volume effects on polymer chains confined to spherical
  surfaces.
\newblock {\em Macromolecules}, 38:1469--1477, 2005.

\bibitem{MFSurfaceInteract}
E.~Katzav, M.~Adda-Bedia, and A.~Boudaoud.
\newblock A statistical approach to close packing of elastic rods and to {DNA}
  packaging in viral capsids.
\newblock {\em Proc. Natl. Acad. Sci}, 103:18900--18904, 2006.

\bibitem{phiStructure}
Y.~Tao, N.~H. Olson, W.~Xu, D.~L. Anderson, M.~G. Rossmann, and T.~S. Baker.
\newblock Assembly of a tailed bacterial virus and its genome release studied
  in three dimensions.
\newblock {\em Cell}, 95:431--437, 1998.

\bibitem{VirExp11}
W.~Ernshaw and S.~Casjens.
\newblock {DNA} packaging by the double-stranded {DNA} bacteriophages.
\newblock {\em Cell}, 21:319--331, 1980.

\bibitem{VirExp2}
W.~Jiang, J.~Chang, J.~Jakana, P.~Weigele, and W.~Chiu.
\newblock Structure of epsilon15 bacteriophage reveals genome organization and
  {DNA} packaging/injection apparatus.
\newblock {\em Nature}, 439:612, 2006.

\bibitem{VirExp1}
J.~Lepault, J.~Dubochet, W.~Baschong, and E.~Kellenberger.
\newblock Organization of double-stranded {DNA} in bacteriophages: A study by
  cryo-electron microscopy of vitrified samples.
\newblock {\em EMBO J.}, 6:1507--1512, 1987.

\bibitem{VirExp3}
M.~Cerritelli, N.~Cheng, A.~H. Rosenberg, C.~E. McPherson, F.~P. Booy, and
  A.~C. Steven.
\newblock Encapsidated conformation of bacteriophage t7 dna.
\newblock {\em Cell}, 91:271, 1997.

\bibitem{VirExp5}
P.~Serwer, S.~Khan, S.~Hayes, R.~Watson, and G.~A. Griess.
\newblock The conformation of packaged bacteriophage t7 dna: Informative images
  of negatively stained t7.
\newblock {\em J. Struct. Biol.}, 120:32, 1997.

\bibitem{VirExp8}
L.~W. Black, W.~W. Newcomb, J.~W. Boring, and J.~C. Brown.
\newblock Ion etching of bacteriophage t4: Support for a spiral-fold model of
  packaged {DNA}.
\newblock {\em Proc. Natl. Acad. Sci.}, 82:7960, 1985.

\bibitem{VirExp10}
L.~W. Black and D.~J. Silverman.
\newblock Model for {DNA} packaging into bacteriophage t4 heads.
\newblock {\em J. Virol.}, 643:643, 1978.

\bibitem{VirExp4}
Z.~Zhang, B.~Greene, P.~A. Thuman-Commike, J.~Janka, P.~Prevelige, J.~King, and
  W.~Chiu.
\newblock Visualization of the maturation transition in bacteriophage p22 by
  electron microscopy.
\newblock {\em J. Mol. Biol.}, 297, 2000.

\bibitem{WidomVirus}
J.~Widom and R.~L. Baldwin.
\newblock Tests of spool models for {DNA} packaging in phage $\lambda$.
\newblock {\em J. Mol. Biol}, 171:419, 1983.

\bibitem{VirExp6}
E.~C. Mendelson, W.~W. Newcomb, and J.~C. Brown.
\newblock Ar$^+$ plasma-induced damage to {DNA} in bacteriophage $\lambda$:
  Implications for the arrangement of {DNA} in the phage head.
\newblock {\em J. Virol.}, 66:2226, 1992.

\bibitem{VirExp7}
S.~L. Novick and J.~D. Baldeshwieler.
\newblock Flouresence measurement of the kinetics of {DNA} injection by
  bacteriophage $\lambda$ into liposomes.
\newblock {\em Biochem.}, 27:7919, 1988.

\bibitem{VirExp9}
J.~C. Brown and W.~W. Newcomb.
\newblock Ion etching of bacteriophage $\lambda$: Evidence that the right end
  of the {DNA} is located at the outside of the phage {DNA} mass.
\newblock {\em J. Virol.}, 60:564, 1986.

\bibitem{HudStruct}
N.~V. Hud.
\newblock Double-stranded dna organization in bacteriophage heads: An
  alternative toroid-based model.
\newblock {\em Biophys. J.}, 69:1355, 1995.

\bibitem{Buste}
S.~Smith, L.~Finzi, and C.~Bustamante.
\newblock Direct mechanical measurements of the elasticity of single {DNA}
  molecules by using magnetic beads.
\newblock {\em Science}, 258:1122--1126, 1992.

\bibitem{HarveyReview}
A.~Petrov and S.~Harvey.
\newblock Packaging double-helical {DNA} into viral capsids: Structures,
  forces, and energetics.
\newblock {\em Biophys J.}, 95:497--502, 2008.

\bibitem{TorroidVirus1}
S.~Tzlil, J.~T. Kindt, W.~M. Gelbart, and A.~Ben-Shaul.
\newblock Forces and pressures in {DNA} packaging and release from viral
  capsids.
\newblock {\em Biophys. J.}, 84:1616--1627, 2003.

\bibitem{TorroidSim}
J.~Kindt, S.~Tzlil, A.~Ben-Shaul, and W.~M. Gelbart.
\newblock {DNA} packaging and ejection forces in bacteriophage.
\newblock {\em Proc. Natl. Acad. Sci}, 98:13671--13674, 2001.

\bibitem{VirusEnergy}
J.~C. LaMarque, T.~L. Le, and S.~C. Harvey.
\newblock Packaging double-helical {DNA} into viral capsids.
\newblock {\em Biopolymers}, 73:348--355, 2003.

\bibitem{MuthVirus}
C.~Forrey and M.~Muthukumar.
\newblock Langevin dynamics simulations of genome packaging in bacteriophage.
\newblock {\em Biophys. J.}, 91:25--41, 2006.

\bibitem{DirectorVirus}
W.~S. Klug and M.~Ortiz.
\newblock A director-field model of {DNA} packaging in viral capsids.
\newblock {\em J. Mec. Phys. Solids}, 51:1815--1847, 2003.

\bibitem{VirusEnergy1}
P.~K. Purohit, J.~Kondev, and R.~Phillips.
\newblock Mechanics of {DNA} packaging in viruses.
\newblock {\em Proc. Natl. Acad. Sci}, 100:3173--3178, 2003.

\bibitem{VirusEnergy2}
P.~K. Purohit, M.~M. Inamdar, P.~D. Grayson, T.~M. Squires, J.~Kondev, and
  R.~Phillips.
\newblock Forces during bacteriophage {DNA} packaging and ejection.
\newblock {\em Biophys. J.}, 88:851--866, 2005.

\bibitem{DiscreteVirus}
P.~K. Purohit, J.~Kondev, and R.~Phillips.
\newblock Force steps during viral {DNA} packaging?
\newblock {\em J. Mech. Phys. Solids}, 51:2239--2257, 2003.

\bibitem{HarveyPhi29}
C.~Locker, S.~Fuller, and S.~Harvey.
\newblock {DNA} organization and thermodynamics during viral packing.
\newblock {\em Biophys J.}, 93:2861--2869, 2007.

\bibitem{OdijkVirus}
T.~Odijk.
\newblock Hexagonally packed {DNA} within bacteriophage t7 stabilized by
  curvature stress.
\newblock {\em Biophys. J.}, 75:1223--1227, 1998.

\bibitem{ViralTC}
D.~Marenduzzo and C.~Micheletti.
\newblock Thermodynamics of {DNA} inside a viral capsid: The role of {DNA}
  intrinsic thickness.
\newblock {\em J. Mo. Biol.}, 330:485--492, 2003.

\bibitem{BloomfieldEnergy}
S.~C. Riemer and V.~A. Bloomfield.
\newblock Packaging of {DNA} in bacteriophage heads: Some considerations on
  energetics.
\newblock {\em Biopolymers}, 17:785, 1978.

\bibitem{HelicalSim}
R.~Metzler and P.~G. Dommersnes.
\newblock Helical packaging of semiflexible polymers in bacteriophages.
\newblock {\em Eur. Biophys. J.}, 33:497, 2004.

\bibitem{ConfScaling}
T.~Sakaue.
\newblock Semiflexible polymer confined in closed spaces.
\newblock {\em Macromol.}, 40:5206, 2007.

\bibitem{EjectionRatchet}
M.~M. Inamdar, W.~M. Gelbart, and R.~Phillips.
\newblock Dynamics of ejection from bacteriophage.
\newblock {\em Biophys. J.}, 91:411--420, 2006.

\bibitem{VirusTwistSim}
A.~J. Spakowizt and Z.~G. Wang.
\newblock {DNA} packing in bacteriophage: Is twist important?
\newblock {\em Biophys. J.}, 88:3192, 2005.

\bibitem{HyeonJCP06}
C.~Hyeon and D.~Thirumalai.
\newblock Kinetics of interior loop formation in semiflexible chains.
\newblock {\em J. Chem. Phys.}, 124:104905, 2006.

\bibitem{MeanField1}
B.~Y. Ha and D.~Thirumalai.
\newblock A mean-field model for semiflexible chains.
\newblock {\em J. Chem. Phys.}, 103:9408, 1995.

\bibitem{MeanFieldForce}
B.~Y. Ha and D.~Thirumalai.
\newblock Semifexible chains under tension.
\newblock {\em J. Chem. Phys.}, 106:4243, 1997.

\bibitem{THBook}
D.~Thirumalai and B.~Y. Ha.
\newblock {\em Theoretical and Mathematical Models in Polymer Research}.
\newblock Acadamia, New York, 1988.

\bibitem{CBDT}
C.~Hyeon and D.~Thirumalai.
\newblock Kinetics of interior loop formation in semiflexible chains.
\newblock {\em J. Chem. Phys.}, 124:104905, 2006.

\bibitem{WLCUnconfInteract}
P.~Hansen and R.~Podgornik.
\newblock Wormlike chains in the large d-limit.
\newblock {\em J. Chem. Phys}, 114:8637, 2001.

\bibitem{Feyn}
R.~P. Feynman and A.~R. Hibbs.
\newblock {\em Quantum Mechanics and Path Integrals}.
\newblock McGraw-Hill, New York, 1965.

\bibitem{CBMCBook}
D.~Frenkel and B.~Smit.
\newblock {\em Understanding Molecular Simulation: From Algorithms to
  Applications.}
\newblock Academic Press, San Diego, 2 edition, 2002.

\bibitem{BusteDNATension}
C.~Bustamante, Z.~Bryant, and S.~B. Smith.
\newblock 10 years of tension: Single-molecule {DNA} mechanics.
\newblock {\em Nature}, 421:423, 2003.

\bibitem{HoneycuttBP92}
J.~D. Honeycutt and D.~Thirumalai.
\newblock The nature of folded states of globular proteins.
\newblock {\em Biopolymers}, 32:695--709, 1992.

\end{thebibliography}
\end{document}